\documentclass[letterpaper,english,aps,preprintnumbers,nofootinbib,showpacs,prd,notitlepage]{revtex4-1}
\usepackage{amsmath}
\usepackage{makeidx}
\usepackage{float}
\usepackage{graphicx}
\usepackage[subrefformat=parens,labelformat=parens,caption=false]{subfig}
\usepackage{cancel}
\usepackage{amsfonts}
\usepackage{xcolor}
\usepackage{diagbox}

\begin{document}

\def\({\left(}
\def\){\right)}
\def\[{\left[}
\def\]{\right]}

\global\long\def\M{\mathcal{M}}
\global\long\def\O{\mathcal{O}}
\global\long\def\GF{\text{G}_\text{F}}
\global\long\def\dd{\text{d}}
\global\long\def\gnn{\text{Ps}\to \gamma \nu_\ell \bar{\nu}_\ell}
\global\long\def\pgnn{\text{p-Ps}\to \gamma \nu_\ell \bar{\nu}_\ell}
\global\long\def\ognn{\text{o-Ps}\to \gamma \nu_\ell \bar{\nu}_\ell}
\global\long\def\pnn{\text{p-Ps}\to \nu_\ell \bar{\nu}_\ell}
\global\long\def\onn{\text{o-Ps}\to \nu_\ell \bar{\nu}_\ell}
\global\long\def\nn{\text{Ps}\to \nu_\ell \bar{\nu}_\ell}
\global\long\def\oggg{\text{o-Ps}\to3\gamma}
\global\long\def\pgg{\text{p-Ps}\to2\gamma}

\preprint{Alberta Thy 19-15}
\title{Positronium Decay into a Photon and Neutrinos}

\author{Andrzej Pokraka}
\email{pokraka@ualberta.ca}

\author{Andrzej Czarnecki}
\email{andrzejc@ualberta.ca}

\affiliation{Department of Physics, University of Alberta, Edmonton, Alberta,
Canada T6G 2E1}

\begin{abstract}
  We determine the rates and energy and angular distributions of the
  positronium decays into a photon and a neutrino-antineutrino pair,
  $\gnn$.  We find that both positronium spin states have access to
  this decay channel, contrary to a previously published result.  The
  low-energy tails of the spectra are shown to be sensitive to the
  binding effects and agree with Low's
  theorem. Additionally, we find a connection between the behaviour of
  the soft photon spectrum in both $\ognn$ and $\oggg$ decays, and the
  Stark effect.
\end{abstract}


\maketitle

\section{\label{sec:intro} Introduction}

Positronium (Ps), the bound state of an electron and its antiparticle,
is a metastable leptonic atom. It is the lightest known atom and in
many ways resembles hydrogen. Like hydrogen, Ps can form two spin
states: the singlet parapositronium (p-Ps) and the triplet
orthopositronium (o-Ps). The lifetimes of Ps are determined by the
electron-positron annihilation rate at rest, $e^+e^-\to 2\gamma$, for
p-Ps \cite{pirenne} and $e^+e^-\to 3\gamma$ for o-Ps
\cite{Ore:1949te}.

Decays of Ps can be precisely described within pure quantum
electrodynamics (QED); the only limitation being the computational
complexity of the higher orders in the expansion in the fine structure
constant $\alpha\simeq 1/137$. Despite this complexity, many
corrections in higher orders have been calculated
\cite{Kniehl:2009pg,Adkins:2015wya,Adkins:2003eh,Adkins:2001zz,Adkins:2000fg,Penin:2014bea,Penin:2003jz,Kniehl:2000dh,Karshenboim:2005iy,Hill:2000qi,Melnikov:2000fi,Czarnecki:1999gv,Czarnecki:1999ci}.

In addition to purely photonic decay modes, weak interactions can
transform Ps into final states involving neutrinos
\cite{Bernreuther:1981ah,Czarnecki:1999mt,Asai:1993np,Maeno:1995gf,Crivelli:2004fr,Badertscher:2006fm}.  
Recently, Ref.~\cite{Perez-Rios:2015lia} examined the exotic decay of Ps into a
photon and a neutrino-antineutrino pair $\gnn$, and claimed that only
p-Ps can decay in this way. On the other hand,
Ref.~\cite{Bernreuther:1981ah} stated that o-Ps can decay into such a
final state and even estimated its branching ratio.

To address the apparent contradiction of \cite{Perez-Rios:2015lia} and
\cite{Bernreuther:1981ah}, we calculate the $\gnn$ decay rates and
photon spectra for both p-Ps and o-Ps (Sec.~\ref{sec:RandS}).  We find
that both p-Ps and o-Ps have access to the $\gnn$ decay mode.
In addition to establishing a non-zero o-Ps rate, we find differences between our calculated
p-Ps rate and spectrum and those of Ref.~\cite{Perez-Rios:2015lia}.
We calculate the angular distributions of $\gnn$ decays in
Sec.~\ref{sec:orthog}. 

It is easy to mislead oneself into thinking that only one Ps spin state
can decay into $\gamma\bar\nu\nu$, since none of the previously studied
final states was accessible to both. 
In pure QED, o-Ps can decay into an odd number of photons and p-Ps into an even
number only, by the charge-conjugation (C) symmetry.
However, the weak bosons couple to both the C-odd
vector and the C-even axial current. Thus, p-Ps can decay into a
photon and a neutrino pair by a vector coupling (analogous to its main
$\gamma\gamma$ decay) while o-Ps can decay into the same final state
through an axial coupling.

In three-body channels, the energy of decay products has an extended
distribution. Its low-energy tail is sensitive to binding effects;
such effects have been determined in the three-photon decay of o-Ps 
\cite{Pestieau:2001ke,Manohar:2003xv,Voloshin:2003hh,RuizFemenia:2007qx}. 
We find analogous phenomena in the 
$\gnn$ decay. In the present case one can compare the low-energy  behaviour of
p-Ps and o-Ps decays, unlike in case of the $3\gamma$ final state, accessible only to
o-Ps. In Sec.~\ref{sec:NREFT} we employ the non-relativistic effective field
theory (NREFT) methods of
\cite{Manohar:2003xv,Voloshin:2003hh,RuizFemenia:2007qx} to explain how binding effects connect the
linear behaviour of the spectra found in Sec.~\ref{sec:RandS} with the
cubic behaviour at extremely low energy, predicted by Low's theorem \cite{Low:1958sn}
(Sec.~\ref{sec:Low}).

\section{\label{sec:RandS} Decay Rates and Spectra}

\begin{figure}[h]
	\centering
   	\subfloat[]{\label{fig:gnnZ1}
  		\includegraphics[width=0.3\textwidth]{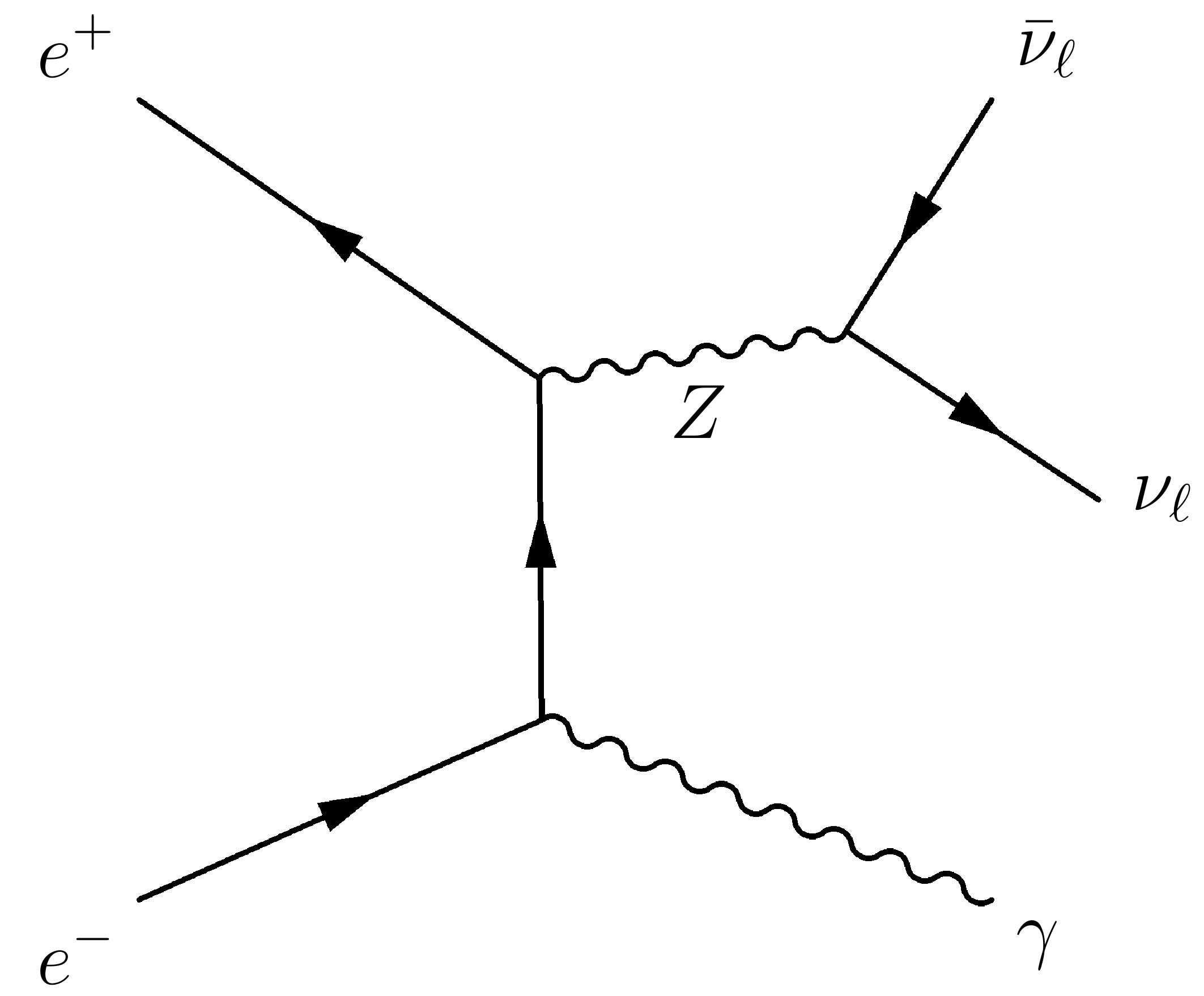}
	}
	\subfloat[]{\label{fig:gnnW1}
  		\includegraphics[width=0.3\textwidth]{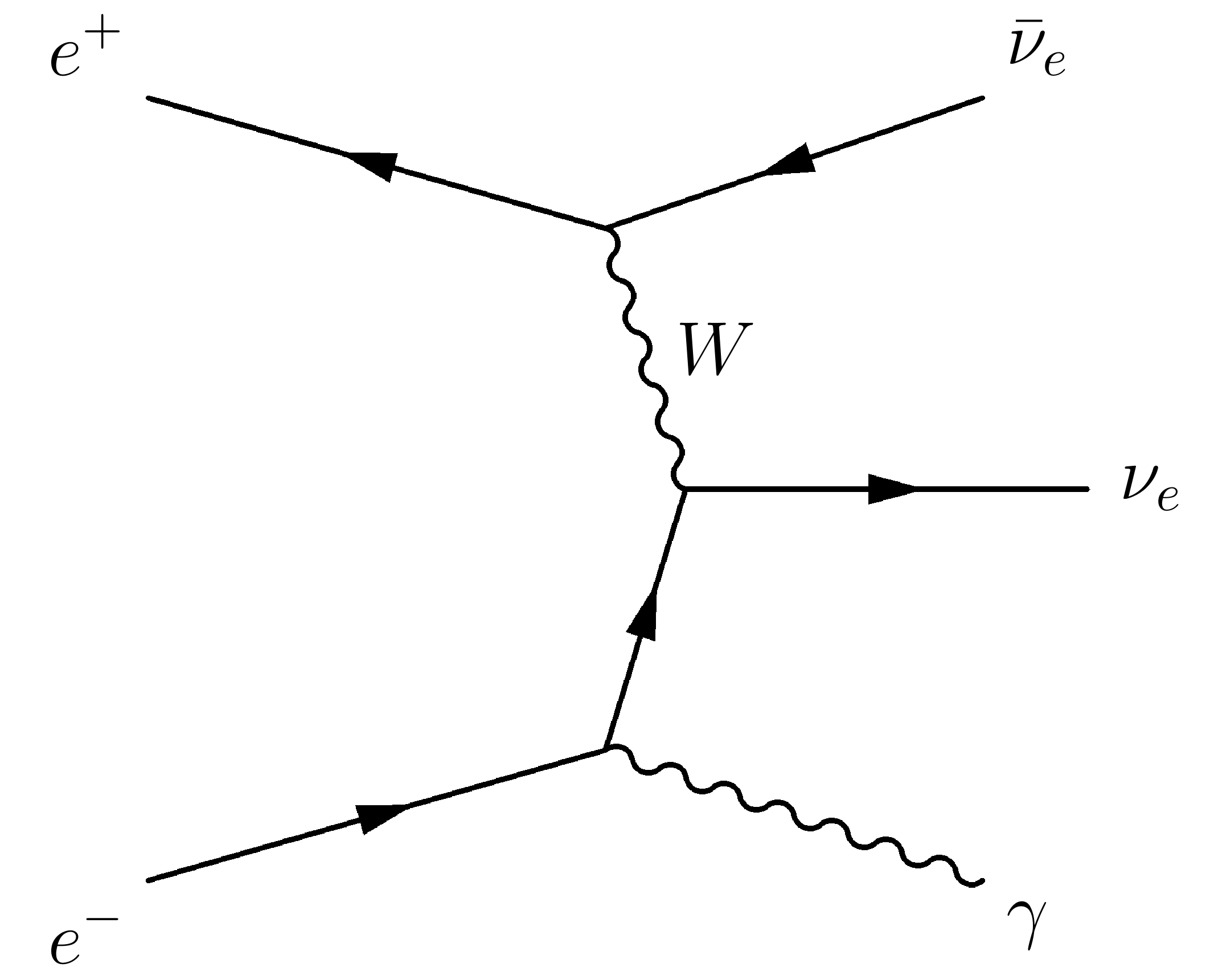}
	}
	\subfloat[]{\label{fig:gnnW3}
  		\includegraphics[width=0.3\textwidth]{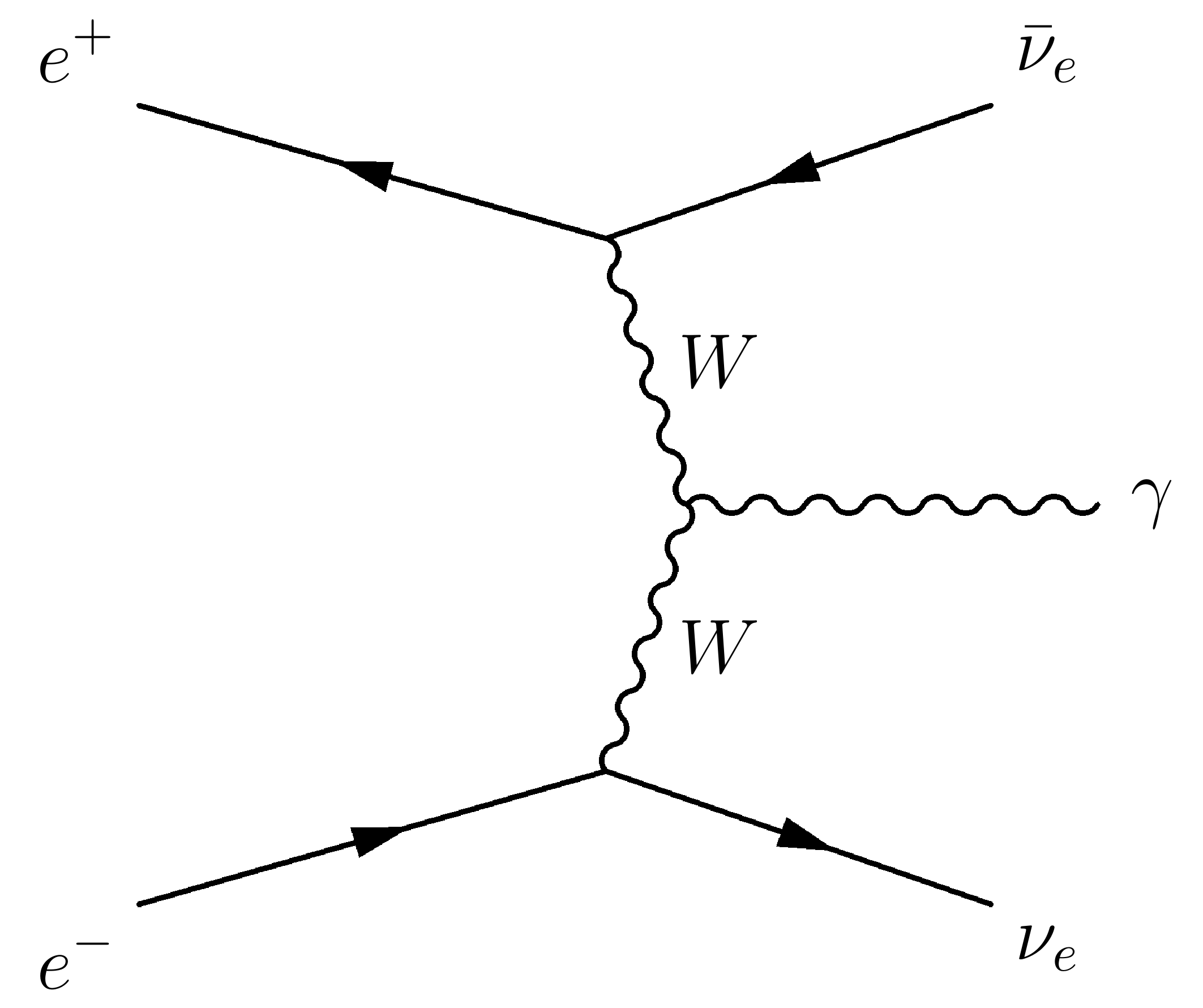}
	}
       \caption{
    		Feynman graphs that contribute to the annihilation
		$e^{+}e^{-}\to\gamma \nu_\ell \bar{\nu}_\ell$ amplitudes 
		relevant for the $\text{Ps}\to\gamma \nu_\ell\bar{\nu}_\ell$
		decays where $\ell=e,\mu,\tau$. For both \protect\subref{fig:gnnZ1} and 
		\protect\subref{fig:gnnW1}, there is an analogous graph where the photon 
		is emitted off the positron leg.  Both \protect\subref{fig:gnnW1} and \protect\subref{fig:gnnW3} only 
		contribute to the amplitude when $\ell=e$.
    		}
	\label{fig:gnn diagrams}
\end{figure}

The relevant $e^+e^-\to\gamma\nu_\ell\bar{\nu}_\ell$ annihilation graphs for $\gnn$ decays are presented in Fig.~\ref{fig:gnn diagrams}. 
The photon is emitted off the initial electron or positron before the $e^+e^-$ pair annihilates into a neutrino-antineutrino pair via $Z$ or $W$ boson exchange. 
The $s$-channel $Z$-boson exchange (Fig.~\protect\subref*{fig:gnnZ1}) contributes to the amplitude for all lepton flavors, $\ell$, while the $t$-channel $W$-boson exchange (Fig.~\protect\subref*{fig:gnnW1}) contributes to the amplitude only when $\ell=e$.  
The photon can also be emitted off of an internal charged $W$ boson (Fig.~\protect\subref*{fig:gnnW3}); since this process is suppressed by an additional factor of $m^2/M_W^2 \ll 1$ where $m$ is the electron mass and $M_W$ is the $W$-boson mass, it is ignored in our calculations. 

We begin by calculating both $\gnn$ decay amplitudes. 
The initial incoming 4-momenta of the electron and positron are denoted by $p_1$ and $p_2$ while outgoing four-momenta are denoted by $k_i$ where $k_1$ is the four-momentum of the neutrino, $k_2$ the anti-neutrino and $k_\gamma$ the photon. 
Since the Ps binding energy is small, $\O(m\alpha^2)$, compared to the rest mass of the initial leptons, their average kinetic energy is negligible. Therefore, we take the initial electron and positron to be at rest with 4-momentum $p_1=p_2=p=(m,\mathbf{0})$. Similarly, the momenta of the virtual $Z$ and $W$ bosons are also negligible compared to their rest masses and their momentum is neglected in the $Z$ and $W$ propagators. To account for the bound state nature of Ps, we include p-Ps and o-Ps projection operators in the spinor trace  of the amplitudes along with a factor of $\psi_0(0)/\sqrt{m}$ where $\psi_0(0)$ is the Ps ground state wavefunction. With these considerations, the $\gnn$ decay amplitudes are
\begin{eqnarray}
	i\mathcal{M}_\text{p/o} 
	&=& -4 \sqrt{2} i e \GF m \frac{\psi_0(0)}{\sqrt{m}}
		 \,\bar{u}(k_1) \gamma_\mu (b_\ell-a_\ell \gamma^5) v(k_2)
	\nonumber \\
	& & \times
		\text{Tr}\,
\Psi_\text{p/o}\(
			\gamma^\mu \left( v_\ell- a_\ell \gamma^5 \right)
				\frac{ \cancel{p}_1-\cancel{k}_\gamma+m }{ (p_1-k_\gamma)^2-m^2 } 
					\cancel{\epsilon}_\gamma^{*}
			+ \cancel{\epsilon}_\gamma^* \frac{ \cancel{k}_\gamma-\cancel{p}_2+m }{ (k_\gamma-p_2)^2-m^2 } 
				\gamma^\mu \left( v_\ell- a_\ell \gamma^5 \)
		\right)
\end{eqnarray}
where $\GF\simeq 1.166 \cdot 10^{-5}/\text{GeV}^2$ is the Fermi
constant \cite{Marciano:1999ih}, $\alpha\simeq 1/137$ is the fine
structure constant, $\epsilon_\gamma$ is the photon polarization and
$\Psi_\text{p/o}$ are the p-Ps and o-Ps projection operators of
Ref.~\cite{Czarnecki:1999mw}. Here, $v_\ell$ and $a_\ell$ describe the
electron vector and axial-vector couplings induced by $Z$ ($\ell
=e,\mu,\tau$) and $W$ ($\ell = e$; a Fierz transformation is
understood \cite{Czarnecki:1999mt}) boson exchange 
\begin{eqnarray}
  	v_\ell &=& 
  	\begin{cases}
		\frac{1}{4} + \sin^2\theta_\text{W} &
  		\mbox{for } \ell= e
		\\
		\frac{1}{4} - \sin^2\theta_\text{W} &
  		\mbox{for } \ell= \mu,\tau,     
  	\end{cases}
	\label{eq:vector couplings}
	\\
  	a_\ell &=& \frac{1}{4} .
	\label{eq:axial couplings}
\end{eqnarray}
Since the weak mixing angle, $\theta_W$, is such that
$\sin^2\theta_\text{W} \simeq 0.238$ \cite{Czarnecki:2005pe}
(numerically close to 1/4), the vector coupling is suppressed for
$\ell = \mu, \tau$. 
We find the total decay rates
\begin{eqnarray}
  	\Gamma_\text{p} \equiv \Gamma(\mbox{p-Ps}  \to \gamma \nu_\ell \bar\nu_\ell)
 	&=& \frac{ 2 \GF^2 m^5 \alpha^4 v_\ell^2}{9\pi^3}
	\approx	\begin{cases}
			3.5 \cdot 10^{-14} \text{ s}^{-1} & \text{for }\ell = e
			\\ 
			2.1 \cdot 10^{-17} \text{ s}^{-1} & \text{for } \ell = \mu,\tau,
		\end{cases}
	\label{eq:p-Ps rate}
	\\
   	\Gamma_\text{o} \equiv \Gamma(\mbox{o-Ps}  \to \gamma  \nu_\ell \bar\nu_\ell )
 	&=& \frac{ 8 \GF^2 m^5 \alpha^4 a_\ell^2}{27\pi^3}
	\approx 1.2 \cdot 10^{-14} \text{ s}^{-1}.
	\label{eq:o-Ps rate}
\end{eqnarray}
The branching ratios are small, as expected for weak decays:
\begin{eqnarray}
 	\text{Br}(\text{p-Ps} \to \gamma \nu \bar{\nu}) 
		\approx \frac{\Gamma (\pgnn)}{\Gamma(\pgg)}
	&\approx& 
 		\begin{cases}
 			4.3 \cdot 10^{-24} & \mbox{for } \ell=e
			\\
			2.6 \cdot 10^{-27} & \mbox{for } \ell=\mu, \tau,
		\end{cases}
	\\
	\text{Br}(\text{o-Ps} \to \gamma \nu \bar{\nu}) 
		\approx \frac{\Gamma (\ognn)}{\Gamma(\oggg)}
	&\approx& 1.7 \cdot 10^{-21}  \mbox{ for } \ell= e,\mu, \tau. 
	\label{eq:o-Ps Br}
\end{eqnarray}
We find that the o-Ps not only can decay radiatively into neutrinos,
but also that since it can decay into all three flavors with equal
probability, its total decay rate into $\nu\bar\nu\gamma$ is in fact
slightly larger than for the p-Ps.

Equation \eqref{eq:o-Ps Br} shows that the o-Ps branching
ratio was overestimated by two orders of magnitude in
\cite{Bernreuther:1981ah}. The estimate of
Ref.~\cite{Bernreuther:1981ah} has the correct powers of the universal
constants, $\GF, \alpha$, and $m$
\begin{equation}
	\frac{\Gamma (\ognn)}{\Gamma(\oggg)}
		\propto \left( \frac{\GF m^2}{\alpha} \right)^2
			\approx 10^{-19}.
\end{equation}
However, the additional factor $4 a_\ell^2 / \left(3\pi^2 (\pi^2-9)
\right) \approx 0.01$ reduces the branching ratio by two orders of
magnitude.

In Ref.~\cite{Perez-Rios:2015lia}, o-Ps is claimed not to decay into
$\gamma \nu \bar{\nu}$, contrary to what we find. On the other hand,
the decay rate of p-Ps into this final state seems to be overestimated
by about a factor 60. Their result, presented as
$\Gamma\(\mbox{p-Ps} \to \gamma \nu_\ell \bar\nu_\ell\)=\frac{\alpha^4
  \GF^2 m^5}{\pi^3}\Sigma(\sin^2\theta_W)$,
has the correct dependence on coupling constants and the mass, but the
function of the weak mixing angle $\Sigma(\sin^2\theta_W)$ seems to be
in error. This can be seen in equation (11) in \cite{Perez-Rios:2015lia} that describes the
decay into  muon neutrinos. Only the $Z$ boson contributes in this
channel, so the amplitude should be proportional to the vector
coupling of the $Z$ to electrons and vanish when
$\sin^2\theta_W\to 1/4$; the expression in that equation does not
vanish in this limit.

\begin{figure}[h] 
	\centering
	\subfloat[]{\label{fig:spectrum a}
       		 \includegraphics[width=.5\textwidth]{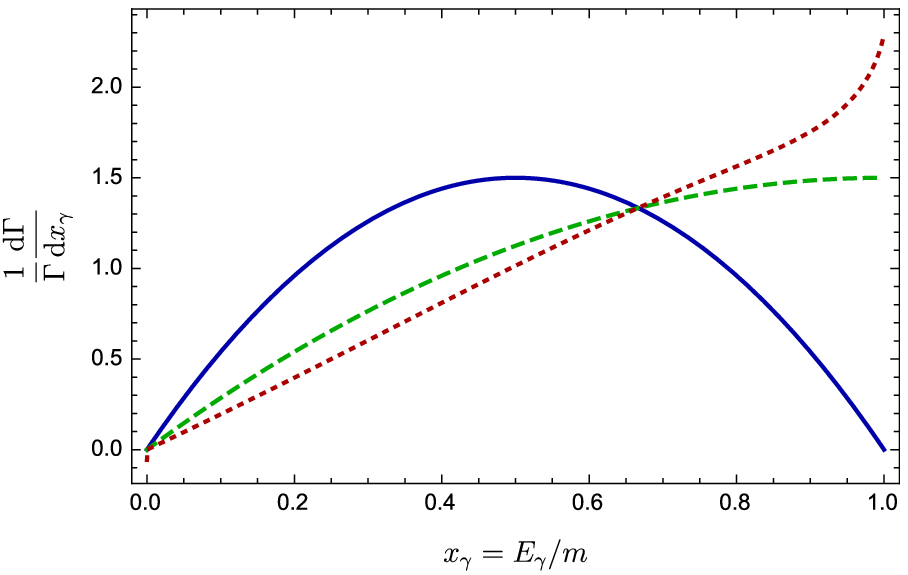}
	}
	\subfloat[]{\label{fig:spectrum b}
		 \includegraphics[width=.5\textwidth]{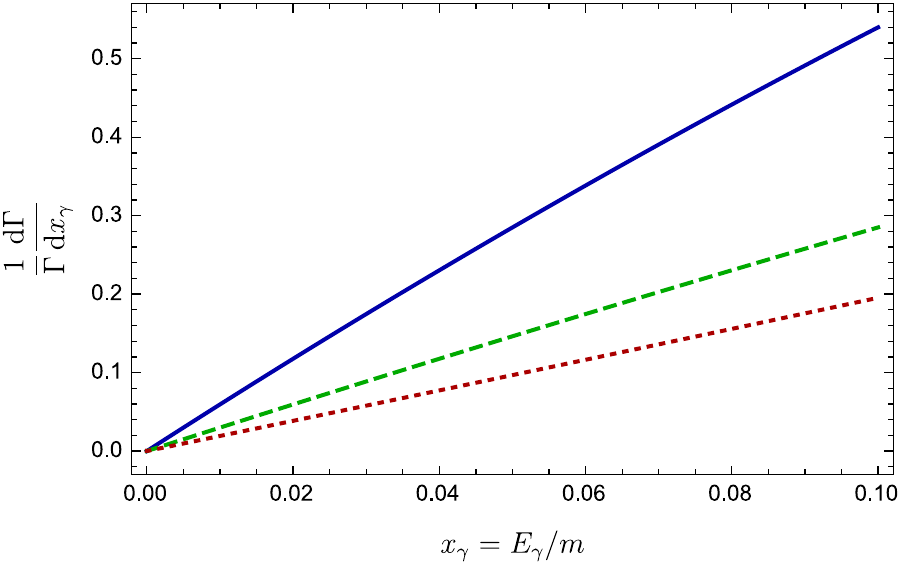}		 
	}
	\caption{
          	 The photon spectrum of $\pgnn$ (solid), $\ognn$ (dashed) and 
		 $\oggg$ (dotted) decays plotted over 
		  \protect\subref{fig:spectrum a} the full energy domain of the photon, $x_\gamma\in(0,1)$
		 and \protect\subref{fig:spectrum b} over the low-energy domain 
		 $x_\gamma\in(0,0.1)$.
	}
	\label{fig:spectrum}
\end{figure}

For the photon spectra we find very simple expressions,
\begin{eqnarray}
  	\frac{1}{\Gamma_\text{p}} \frac{\dd\Gamma_\text{p}}{\dd x_\gamma} 
		&=&  6 x_\gamma (1-x_\gamma), 
	\label{eq:p-Ps spectrum}
	\\
  	\frac{1}{\Gamma_\text{o}} \frac{\dd\Gamma_\text{o}}{\dd x_\gamma} 
		&=&  \frac{3}{2} x_\gamma (2-x_\gamma),
	\label{eq:o-Ps spectrum}
\end{eqnarray}
where $x_\gamma=E_\gamma/m \in (0,1)$.  These spectra are shown in
Fig.~\ref{fig:spectrum}. Since there is some similarity between $\gnn$
and $\oggg$ decays, the $\oggg$ spectrum (first calculated by Ore and
Powell \cite{Ore:1949te}) is also included in Fig.~\ref{fig:spectrum}
for comparison.
When the photon reaches the maximum energy, $x_\gamma=1$, the neutrino
(left-handed) and the antineutrino (right-handed) move collinearly in
the direction opposite to the photon. Their spins cancel and the
angular momentum of the system is carried by the photon's
spin. Clearly, this is possible only for o-Ps; for this reason, the
p-Ps spectrum vanishes at $x_\gamma=1$
(Fig.~\protect\subref*{fig:spectrum a}). This spectrum also vanishes
at $x_\gamma=0$. However, the p-Ps spectrum of
Ref.~\cite{Perez-Rios:2015lia} vanishes at neither $x_\gamma=0$ or
$x_\gamma=1$.

The p-Ps spectrum is maximal at $x_\gamma=1/2$; different from the
maximum $x_\gamma=2/3$, predicted in \cite{Perez-Rios:2015lia}. On the
other hand the o-Ps spectrum is maximal at $x_\gamma=1$ when the
photon carries whole angular momentum of the system.

We also note that the spectra we have found (neglecting binding
effects) are linear in the low-energy limit
(Fig.~\protect\subref*{fig:spectrum b}). Since Low's theorem
\cite{Low:1958sn} predicts the low-energy behaviour of the spectrum to
be cubic rather than linear, we shall determine how binding effects
modify the results \eqref{eq:p-Ps spectrum} and \eqref{eq:o-Ps spectrum} (Sec.~\ref{sec:Low}).

\section{\label{sec:orthog} Angular Distributions of $\boldsymbol{\gnn}$ Decays}

In Sec.~\ref{sec:RandS}, we calculated the decay rates and spectra for p-Ps and o-Ps, and found that both  can decay into a photon and a neutrino-antineutrino pair. 
To better understand these decays, we calculate the angular dependence of the $\gnn$ amplitudes (Sec.~\ref{sec:ang amp}) and then use those amplitudes to determine the angular distributions of $\gnn$ decays (Sec.~\ref{sec:ang dis}).

\subsection{ \label{sec:ang amp} Angular Dependence of the Decay Amplitudes}
The angular dependence of the decay amplitudes is most easily found by
reformulating the three-body decay $\gnn$, in terms of a two-body
decay $\text{Ps}\to\gamma Z^*$, where $Z^*$ is a massive vector boson
of polarization $\epsilon$ and 4-momentum $q$. Specifically, the
three-body phase space of the decay rate is factorized into two
two-body phase spaces (one for $\text{Ps}\to\gamma Z^*$ and one for
$Z^*\to\nu\bar{\nu}$) and an integral over the invariant mass squared
of the $Z^*$ boson. 
After integrating over the neutrino momenta, the $\gnn$ decay rate can be written as the integral of the $\text{Ps}\to\gamma Z^*$ decay rate (multiplied by a factor from the $Z^*\to\nu\bar{\nu}$ phase space) over the invariant mass of $Z^*$ squared (Appendix \hyperref[sec:appA]{A}),
\begin{equation}
	\Gamma_\text{p/o}  
	=  \frac{\GF^2}{2\pi^2 \alpha}\int\frac{\dd q^{2}}{2\pi}q^{2}
	\Gamma_{\text{(p/o)-Ps}\to\gamma Z^*},
\label{eq:DR in terms of Z*}
\end{equation}
where $q=k_1+k_2$ is the $Z^*$ 4-momentum and
\begin{eqnarray}
	\Gamma_{\text{(p/o)-Ps} \to \gamma Z^{*} } 
	& = & \frac{1}{g} \frac{1}{2m_{ \text{Ps} } } 
		\int \dd \Phi_2 \left(p_1+p_2;q, k_\gamma \right) 
			\frac{ \left| \psi_{0}(0) \right|^{2} }{m} \frac{1}{3} 
				\sum_\text{spin/pol} \left| \M_{\text{(p/o)-Ps} \to \gamma Z^*} \right|^2.
	\label{eq:gammaZ decay rate}
\end{eqnarray}
Here, $g$ is the number of polarizations of the initial Ps state.

From (\ref{eq:DR in terms of Z*}), it is clear that the three-body problem $\gnn$ can be described in terms of the two-body problem $\text{Ps}\to\gamma Z^*$. The $Z^{*}$ couples to the electron current through both vector and axial-vector coupling with the Feynman rule $i e \cancel{\epsilon}^* \left( v_\ell - a_\ell \gamma^5 \right)$ at each $e^\pm Z^*$ vertex. 

To construct the angular dependence of the $\text{Ps} \to \gamma Z^*$ decay amplitudes on the spherical angles $\theta$ and $\phi$, we first determine the decay amplitudes to final states where the photon moves along the $+z$-axis and the $Z^*$ boson moves along the $-z$-axis. The angular dependence is then determined by rotating the initial state and considering decay along the new $z^\prime$-axis \cite{Feynman:1963uxa}. Alternatively, one can obtain the angular dependence using the helicity basis formalism of Refs.~\cite{Jacob:1959at,Richman:1984gh,Leader:2001gr,Chung:1971ri}.

The $\text{p-Ps}\to\gamma Z^*$ decay amplitudes are isotropic and given by
\begin{equation}
	\mathcal{A}_{m^\prime_\gamma,m^\prime_Z}(\theta,\phi)
		=\mathcal{A}_{m^\prime_\gamma,m^\prime_Z}
			\propto \delta_{m^\prime_{\gamma},-m^\prime_{Z}},
	\label{eq:p-Ps ang amp}
\end{equation}
where  $m_{\gamma}^{\prime}\in\{\pm1\}$ and $m_{Z}^{\prime}\in\{\pm1,0\}$ are the spin projections of the photon and $Z^{*}$ along the $z^\prime$-axis. The $z^\prime$-axis points along the photon trajectory defined by the spherical polar angles $\theta$ and $\phi$ in the original unrotated frame. The p-Ps amplitudes were calculated and are listed in Table \ref{tab:p-Ps amp}.

\begin{table}[h]
\caption{ 
	\label{tab:p-Ps amp}
	The  $\text{p-Ps}\to\gamma Z^{*}$  decay amplitudes, 
	$\mathcal{A}_{m_{\gamma}^{\prime}m_{Z}^{\prime}}/v_\ell e^{2}$, 
	as a function of the spherical angles $\theta$ and $\phi$. 
	Since p-Ps is odd under parity, 
	$\mathcal{A}_{-m_\gamma^\prime -m_Z^\prime}=-\mathcal{A}_{m_\gamma^\prime m_Z^\prime}$;
	therefore, only the $m_\gamma^\prime=+1$ amplitudes need be tabulated.
}
\begin{ruledtabular}
	\begin{tabular}{ccccc}
	\diagbox[width=5em]{$m_\gamma^\prime$}{$m_Z^\prime$} 
	& $+1$ 
	& $0$ 
	& $-1$
	\\
	\hline
	$+1$ 
	& $0$
	& $0$
	& $4i/\sqrt{2}$
	\end{tabular}
\end{ruledtabular}
\end{table}

The $\text{o-Ps}\to\gamma Z^{*}$ amplitudes must be calculated for each initial polarization of o-Ps. 
In the initial frame before decay, the o-Ps atom is in a state of definite angular momentum with some spin projection along the $z$-axis. We let $|\Lambda\rangle$ represent this initial state. The o-Ps atom subsequently decays along the $z^\prime$-axis with the amplitude $\mathcal{A}_{m_{\gamma}^{\prime}m_{Z}^{\prime}}^{m_{\Lambda}}$ where $m_{\Lambda}\in\{\pm1,0\}$ is the initial spin projection of o-Ps along the $z$-axis. The o-Ps amplitudes are derived in Appendix \hyperref[appB]{B} and are listed in Table \ref{tab:o-Ps amp}.

\begin{table}[h]
	\caption{ 
	\label{tab:o-Ps amp}
	The $\text{o-Ps}\to\gamma Z^{*}$ decay amplitudes,
	$\mathcal{A}_{m_{\gamma}^{\prime}m_{Z}^{\prime}}^{m_{\Lambda}}/a_\ell e^{2}$,
	 as a function of the spherical angles $\theta$ and $\phi$ evaluated at
	 $\mathbf{q}=-\mathbf{k}_\gamma,E_{Z}=2m-E_\gamma$.
	  The $m_\Lambda = -1$ amplitudes can be obtained from the 
	 $m_\Lambda = +1$ amplitudes
	 by the replacment $\theta\to\theta+\pi$ and $\phi\to-\phi$.	
	}
	\begin{ruledtabular}
	\begin{tabular}{ccccc}
		$m_\Lambda$
	 	& \diagbox[width=5em]{$m_\gamma^\prime$}{$m_Z^\prime$} 
	 	& $+1$ 
	 	& $0$ 
	 	& $-1$
		\\  
		\hline
		$+1$ 
		& $+1$ 
		& 0 
		& $\sqrt{2}(1+\cos\theta)e^{i\phi}/\sqrt{1-x_\gamma}$ 
		& $-2i\sin\theta e^{i\phi}$ 
		\\
	 	& $-1$ 
 		& $2i\sin\theta e^{i\phi}$ 
		& $-\sqrt{2}(1-\cos\theta)e^{i\phi}/\sqrt{1-x_\gamma}$ 
 		& 0 
		\\
		$0$ 
		& $+1$ 
		& 0 
 		& $2\sin\theta/\sqrt{1-x_\gamma}$ 
		& $4i\cos\theta/\sqrt{2}$ 
		\\
		& $-1$ 
		& $-4i\cos\theta/\sqrt{2}$ 
 		& $-2\sin\theta/\sqrt{1-x_\gamma}$ 
 		& 0 
	\end{tabular}
\end{ruledtabular}
\end{table}

To validate the amplitudes in Tables \ref{tab:p-Ps amp} and \ref{tab:o-Ps amp}, we use them to calculate the decay rates and photon spectra, and compare these with those obtained in Sec.~\ref{sec:RandS}. To do this, we first derive the spin averaged amplitudes squared. For p-Ps, this task is simple, 
\begin{equation}
	\frac{1}{3}\sum_{m_{\gamma}^{\prime}m_{Z}^{\prime}}
		\left|
			\mathcal{A}_{m_{\gamma}m_{Z}^{\prime}}
		\right|_{\mathbf{q}=-\mathbf{k}_\gamma,E_{Z}=2m-E_\gamma}^{2}
	= \frac{16v_{\ell}^{2}e^{4}}{3}.
\label{eq:p-Ps spin averaged Asqrd}
\end{equation}
To obtain the o-Ps spin averaged amplitude squared, it is convenient to first  sum over $m_{\Lambda}$ and $m_{\gamma}^{\prime}$
\begin{eqnarray}
	\sum_{m_{\Lambda}m_{\gamma}^{\prime}}
		\left|
			\mathcal{A}_{m_{\gamma}+}^{m_{\Lambda}}
		\right|_{\mathbf{q}=-\mathbf{k}_\gamma,E_{Z}=2m-E_\gamma}^{2}
	&=& \sum_{m_{\Lambda}m_{\gamma}^{\prime}}
		\left|
			\mathcal{A}_{m_{\gamma}-}^{m_{\Lambda}}
		\right|_{\mathbf{q}=-\mathbf{k}_\gamma,E_{Z}=2m-E_\gamma}^{2} 
	= 8a_{\ell}^{2}e^{4},
	\\
	\sum_{m_{\Lambda}m_{\gamma}^{\prime}}
		\left|
			\mathcal{A}_{m_{\gamma}0}^{m_{\Lambda}}
		\right|_{\mathbf{q} = -\mathbf{k}_\gamma,E_{Z}=2m-E_\gamma}^{2}
	&=& \frac{16a_{\ell}^{2}e^{4}}{m-E_\gamma}.
\end{eqnarray}
Then completing the sum over $m_{Z}^{\prime}$ and dividing by the number of o-Ps and $Z^*$ polarizations yields the spin averaged amplitude squared
\begin{equation}
	\frac{1}{3\cdot3} \sum_{m_{\Lambda}m_{\gamma}^{\prime}m_{Z}^{\prime}}
		\left|
			\mathcal{A}_{m_{\gamma}m_{Z}^{\prime}}^{m_{\Lambda}}
		\right|_{\mathbf{q}=-\mathbf{k}_\gamma,E_{Z}=2m-E_\gamma}^{2}
	= \frac{16a_{\ell}^{2}e^{4}}{9}\frac{2-x_\gamma}{1-x_\gamma}.
\label{eq:o-Ps spin averaged Asqrd}
\end{equation}

The decay rates and spectra are calculated by substituting equations \eqref{eq:p-Ps spin averaged Asqrd} and \eqref{eq:o-Ps spin averaged Asqrd} into \eqref{eq:DR in terms of Z*}. Since the spin averaged amplitudes squared are independent of $\theta$ and $\phi$, the angular integrations of \eqref{eq:DR in terms of Z*} are easy and  yield
\begin{eqnarray}
	\Gamma_\text{p/o} 
		&=& \frac{4 \GF^2 m^5 \alpha^4 }{3\pi^3}
	 	\int_0^1 \dd x \; 
		\left\{
			\begin{matrix}
   				 v_\ell^2  x_\gamma (1-x_\gamma)    	\\
  				 a_\ell^2  x_\gamma (2-x_\gamma) / 3 	\\
   			\end{matrix} 
		\right\}
		\label{eq:spectra via Z*}
		\\
		&=& \frac{4 \GF^2 \alpha^4 m^5}{9\pi^3}
		\left\{
			\begin{matrix}
   				 v_\ell^2 / 2   	\\
  				 2 a_\ell^2 / 3	\\
   			\end{matrix} 
		\right\},
		\label{eq:DR via Z*}
\end{eqnarray}
where the top (bottom) line in the curly brackets is used for the p-Ps (o-Ps) decay rate. 
The decay rates \eqref{eq:DR via Z*} are identical to \eqref{eq:p-Ps rate} and \eqref{eq:o-Ps rate}. The spectra are the integrands of equation \eqref{eq:spectra via Z*} and are also equal to the spectra \eqref{eq:p-Ps spectrum} and \eqref{eq:o-Ps spectrum}. Thus, the amplitudes of Tables \ref{tab:p-Ps amp} and \ref{tab:o-Ps amp} are consistent with our results from Sec.~\ref{sec:RandS}.

While it is evident that p-Ps and o-Ps cannot decay into the same final states (even though they have the same constituent particles), we confirm the orthogonality of the p-Ps and o-Ps decay amplitudes.
The o-Ps amplitudes, $\mathcal{A}^{m_\Lambda}_{\pm^\prime 0^\prime}$,  are trivially orthogonal to the p-Ps amplitudes (\ref{eq:p-Ps ang amp}) because p-Ps cannot decay into a longitudinally polarized $Z^*$ and photon. To check the orthogonality of $\mathcal{A}^{m_\Lambda}_{\pm^\prime \mp^\prime}$  with (\ref{eq:p-Ps ang amp}), we take their inner product
\begin{equation}
	\int \dd\Omega\mathcal{A}_{m_\gamma,m_Z} \left(\mathcal{A}_{\pm^{\prime}\mp^{\prime}}^{m_{\Lambda}}\left(\theta,\phi\right) \right)^*
	\propto
	\int \dd\Omega \left( \mathcal{A}_{\pm^{\prime}\mp^{\prime}}^{m_{\Lambda}}\left(\theta,\phi\right)\right)^*.
\end{equation}
Since $\mathcal{A}_{\pm^{\prime}\mp^{\prime}}^{m_{\Lambda}}\left(\theta,\phi\right)$ are proportional to $e^{\pm i\phi}$ or $\cos\theta$ (depending on $m_\Lambda$), the inner products vanish proving orthogonality; this is as expected because $\mathcal{A}_{\pm^{\prime}\mp^{\prime}}^{m_{\Lambda}}\left(\theta,\phi\right)$ (Table~\ref{tab:o-Ps amp}) are p-waves while the p-Ps amplitudes are s-waves (Table~\ref{tab:p-Ps amp}).

Thus, the $\pgnn$ and $\ognn$ decays do not have access to the same final state despite the fact that the final states contain the same constituent particles.

\subsection{\label{sec:ang dis} Angular Distributions}

The angular distribution for a specific $\gamma + Z^*$ final state is found by differentiating the decay amplitude \eqref{eq:DR in terms of Z*}, where the squared amplitude corresponding to the specific final state (Tables \ref{tab:p-Ps amp} and \ref{tab:o-Ps amp}) is used in place of the spin averaged amplitude squared, by $x_\gamma$ and $\cos\theta$. 

Since the p-Ps amplitudes are isotropic, the $\pgnn$ angular distributions are also isotropic ($9 \Gamma_\text{p} x_\gamma (1-x_\gamma) / 2 $ for $m_\gamma^\prime =\pm 1$). Thus, p-Ps is equally likely to decay into a photon and a neutrino-antineutrino pair where the photon is emitted in any direction.

The $\ognn$ angular distributions are determined to be
\begin{equation}
	\frac{1}{\Gamma_\text{o}} \frac{\dd^2 \Gamma_{m_\gamma^\prime m_Z^\prime}^{m_\Lambda}}{\dd x_\gamma \dd\cos\theta} 
	= \frac{27}{64} x_\gamma (1-x_\gamma)
	\int \frac{\dd \phi}{2\pi}
	  \left|
		\frac{\mathcal{A}_{m_\gamma^\prime m_Z^\prime}^{m_\Lambda}}{a_\ell e^2}
	\right|_{\mathbf{q}=-\mathbf{k}_\gamma,E_{Z}=2m-E_\gamma}^{2},
\end{equation}
and are tabulated in Table \ref{tab:ang dis}. 
Since $Z^*$ is a mathematical convenience, the physical angular distributions for a given o-Ps polarization $m_\Lambda$ and photon helicity $m_\gamma^\prime$ is obtained by averaging over the $Z^*$ polarizations. 
For an o-Ps atom initially polarized in the $m_\Lambda=0$ state, the angular distributions for decay into a photon of helicity $m_\gamma^\prime \pm 1$ are $9 \Gamma_\text{o} (\sin^2\theta x_\gamma+2\cos^2\theta x_\gamma(1-x_\gamma))/16$ and non-zero for all $\theta$. 
The angular distribution for o-Ps initially polarized in the $m_\Lambda = +1$ state decaying into a photon of helicity $m_\gamma^\prime = +1$ is $9\Gamma_\text{o}(2\cos^4(\theta/2)x_\gamma + \sin^2\theta x_\gamma (1-x_\gamma))/16$; 
since this angular distribution vanishes for $\theta = \pi$, an o-Ps atom in the $m_\Lambda = +1$ state cannot decay into a photon of helicity $m_\gamma^\prime = +1$ along the $-z$-axis. 
Similarly, an o-Ps atom initially polarized in the $m_\Lambda = +1$ state cannot decay into a photon of helicity $m_\gamma^\prime =-1$ along the $+z$-axis. 

\begin{table}[h]
	\caption{ 
	\label{tab:ang dis}
	The angular distributions for $\ognn$ decays, $(\dd^2 \Gamma_{m_\gamma^\prime m_Z^\prime}^{m_\Lambda}/\dd x_\gamma \dd\cos\theta)/\Gamma_\text{o}$.
	 The $m_\Lambda = -1$ distributions can be obtained 
	 from the $m_\Lambda = +1$ angular distributions
	 by the replacement $\theta\to\theta+\pi$ and $\phi\to-\phi$.
	}
	\begin{ruledtabular}
	\begin{tabular}{ccccc}
		$m_\Lambda$
	 	& \diagbox[width=5em]{$m_\gamma^\prime$}{$m_Z^\prime$} 
	 	& $+1$ 
	 	& $0$ 
	 	& $-1$
		\\  
		\hline
		$+1$ 
		& $+1$ 
		& 0 
		& $27 \cos^4(\theta/2) x_\gamma /8$ 
		& $27 \sin^2\theta x_\gamma (1-x_\gamma)/16$ 
		\\
	 	& $-1$ 
 		& $27 \sin^2\theta x_\gamma (1-x_\gamma)/16$ 
		& $27 \sin^4(\theta/2) x_\gamma/8$ 
 		& 0 
		\\
		$0$ 
		& $+1$ 
		& 0 
 		& $27\sin^2\theta x_\gamma/16$ 
		& $27 \cos^2\theta x_\gamma (1-x_\gamma)/8$ 
		\\
		& $-1$ 
		& $27\cos^2\theta x_\gamma (1-x_\gamma)/8$ 
 		& $27 \sin^2\theta x_\gamma/16$ 
 		& 0 
	\end{tabular}
\end{ruledtabular}
\end{table}

 The photon spectrum for a specific $\gamma + Z^*$ final state is calculated by integrating the corresponding angular distribution by $\dd \cos\theta$. 
These spectra are listed in Tables \ref{tab:p-Ps partial DR} and \ref{tab:o-Ps partial DR} and
provides further insight into equations \eqref{eq:p-Ps spectrum} and \eqref{eq:o-Ps spectrum}.

\begin{table}[h]
\caption{ 
	\label{tab:p-Ps partial DR}
	$\pgnn$ photon spectra, $(\dd \Gamma_\text{p}/\dd x_\gamma) / \Gamma_\text{p}$, 
	for specific $\gamma+Z^{*}$ final states. 
	Since $|\mathcal{A}_{m_\gamma^\prime m_Z^\prime}|^2 
	= |\mathcal{A}_{-m_\gamma^\prime -m_Z^\prime}|^2$, 
	only the $m_\gamma^\prime = +1$ decay rates need be tabulated.
	}
\begin{ruledtabular}
	\begin{tabular}{cccc}
	\diagbox[width=5em]{$m_\gamma^\prime$}{$m_Z^\prime$} 
	& $+1$ 
	& $0$ 
	& $-1$
	\\
	\hline
	$+1$ 
	& $0$
	& $0$ 
	& $9 x_\gamma (1-x_\gamma)$
	\end{tabular}
\end{ruledtabular}
\end{table}

\begin{table}[h]
	\caption{ 
		\label{tab:o-Ps partial DR}
		$\ognn$ photon spectra, $(\dd \Gamma_\text{o}/ \dd x_\gamma) / \Gamma_\text{o}$, 
		for specific $\gamma+Z^{*}$ final states and any $m_\Lambda$.
	}
	\begin{ruledtabular}
	\begin{tabular}{cccc}
	 	 \diagbox[width=5em]{$m_\gamma^\prime$}{$m_Z^\prime$} 
	 	& $+1$ 
	 	& $0$ 
	 	& $-1$
		\\  
		\hline
		 $+1$ 
		& 0 
		& $9 x_\gamma$ /4
		& $9 x_\gamma(1-x_\gamma) /4$ 
		\\
		$-1$ 
 		& $9 x_\gamma(1-x_\gamma) /4$ 
		& $9 x_\gamma /4$ 
 		& 0 
		\\
	\end{tabular}
\end{ruledtabular}
\end{table}

The photon spectrum of decays to final states with $m_{\gamma}^{\prime}=\pm$ and $m_Z^\prime=\mp$ are proportional to $ x_\gamma\left(1-x_\gamma\right)$ and vanish as $x_\gamma \to 1$. On the other hand, the photon spectrum of decays to final states with $m_{\gamma}^{\prime}=\pm$ and $m_Z^\prime=0$ are linear and maximal at $x_\gamma\to1$.  The o-Ps photon spectrum is maximal at $x_\gamma=1$ because the o-Ps decay has access to two additional final states with a longitudinally polarized $Z^*$; these add to the linear term in the spectrum. The $\mathcal{A}^{m_\Lambda}_{\pm0}$ amplitudes contain a factor of $2m/q=1/\sqrt{1-x_\gamma}$ from the longitudinal polarization of $Z^*$.  This factor enhances the amplitude for high-energy photons and cancels the factor $q^{2}\propto\left(1-x_\gamma\right)$ in the $\dd q^{2}$ integral of \eqref{eq:DR in terms of Z*}.  In the high-energy limit, $x_\gamma \to 1$, the longitudinal polarization of $Z^*$ represents  a final state where the neutrino and antineutrino are collinear.

\section{\label{sec:Low} Low's theorem and the Soft Photon Limit of the Spectra}

Low's theorem \cite{Low:1958sn} places constraints on the amplitude of
any radiative process and predicts the spectrum in the soft photon
limit. In Sec.~\ref{sec:RandS}, the tree level electroweak photon
spectra, equations \eqref{eq:p-Ps spectrum} and \eqref{eq:o-Ps
  spectrum}, were found to be linear in the low-energy limit, similar
to the Ore-Powell $\oggg$ spectrum. However, it was pointed out by
Ref.~\cite{Pestieau:2001ke} that the Ore-Powell spectrum is in
contradiction with Low's theorem. Therefore, it is important to
reconcile equations \eqref{eq:p-Ps spectrum} and \eqref{eq:o-Ps
  spectrum} with Low's theorem.

Low's theorem states that the $\mathcal{O}(E^{-1}_\gamma)$ and $\mathcal{O}(E^0_\gamma)$ terms in the Laurent expansion of the  radiative amplitude, $X\to Y+\gamma$, are obtained from knowledge of the non-radiative amplitude, $X\to Y$  \cite{Manohar:2003xv,Low:1958sn,Pestieau:2001ke}.  Expanding the radiative amplitude, $\epsilon_\gamma^{\mu}\M_{\mu}$, in a Laurent series in the photon energy, we obtain
\begin{equation}
	\epsilon_\gamma^{\mu}\M_{\mu}=\sum_{n=-1}^\infty \M_{n}E^{n}_\gamma, 
\end{equation}
where 
$\M_{i}$ is the coefficient of the $\mathcal{O}\left(E^{i-1}_\gamma \right)$ term of the Laurent series. The coefficients $\M_{0}$ and $\M_{1}$ are independent of $E_\gamma$ and determined by the non-radiative amplitude, its derivatives in physically allowed regions and the anomalous magnetic moments of the particles involved in the reaction \cite{Low:1958sn}.

The $\M_{0}$ coefficient is proportional to the non-radiative amplitude multiplied by the factor $-Q_{i}\epsilon\cdot p_{i}/k_\gamma \cdot p_{i}$, which arises from the emission of a photon by an outgoing or ingoing particle \cite{Manohar:2003xv}. The $\M_{0}$ coefficient vanishes when there are no moving charged particles in the initial and final state of the non-radiative process or when the non-radiative amplitude is zero. The coefficient $\M_{1}$ is a function of the magnetic moments of the particles as well as the non-radiative amplitude and its derivatives with respect energy and angle \cite{Low:1958sn}.

By combining the behavior of the radiative amplitude and the phase space, we find that the low-energy photon spectrum has the form 
\begin{equation}
	\frac{\dd\Gamma}{\dd E_{\gamma}}
	= \frac{A}{E_{\gamma}}+B+CE_{\gamma}+DE_{\gamma}^{2}+\mathcal{O}\left(E_{\gamma}^{3}\right),
\end{equation}
where 
\begin{eqnarray}
	A & = & \vert\M_{0}\vert^{2}
	\nonumber \\
	B & = & \M_{0}\M_{1}^{*}+\M_{1}\M_{0}^{*}
	\nonumber \\
	C & = & \left|\M_{1}\right|^{2}+\M_{0}\M_{2}^{*}+\M_{2}\M_{0}^{*}
	\nonumber \\
	D & = & \M_{0}\M_\gamma^{*}+\M_\gamma\M_{0}^{*}+\M_{1}\M_{2}^{*}+\M_{2}\M_{1}^{*}.
\end{eqnarray}
If $\M_{0}$ vanishes, then $A=B=0$ and the soft photon spectrum is of order $E_{\gamma} \dd E_{\gamma}$. If both $\M_{0}$ and $\M_{1}$ vanish, then $A=B=C=D=0$ and the soft photon spectrum is of order $E_{\gamma}^{3} \dd E_{\gamma}$.

For $\pgnn$, the non-radiative $\pnn$ amplitude vanishes \cite{Czarnecki:1999mt};  application of Low's theorem yields $\M_{0,1}=0$ for the radiative decay, $\pgnn$. Since the radiative $\ognn$ decay proceeds only via axial-vector coupling while the non-radiative $\onn$ amplitude is proportional to vector coupling  \cite{Czarnecki:1999mt}, Low's theorem requires that the $\mathcal{O}(E^{-1}_\gamma)$ and $\mathcal{O}(E^0_\gamma)$ terms of the  radiative $\ognn$ amplitude vanish $(i.e., \M_{0,1}=0$). Thus, for both $\gnn$  decays, Low's theorem predicts that the photon spectra are cubic in the low-energy limit in apparent contradiction with equations \eqref{eq:p-Ps spectrum} and \eqref{eq:o-Ps spectrum}.

Equations \eqref{eq:p-Ps spectrum} and \eqref{eq:o-Ps spectrum} were calculated using the tree level electroweak amplitude for the $e^+e^-\to\gamma \nu_\ell \bar{\nu}_\ell$ annihilation multiplied by the probability density for the $e^+e^-$ pair to be at the origin. This calculation assumes that the electron and positron are initially free and at rest, and therefore neglects the binding effects in Ps. Binding effects are of order $m\alpha^2$. For photons with comparable energies, binding effects become important and equations \eqref{eq:p-Ps spectrum} and \eqref{eq:o-Ps spectrum} are no longer accurate.

To resolve the contradiction between equations \eqref{eq:p-Ps spectrum} and \eqref{eq:o-Ps spectrum}, and Low's theorem, we must include binding effects in the soft photon spectrum of $\gnn$ decays. To do this we employ the NREFT methods developed in Refs.~\cite{Manohar:2003xv,RuizFemenia:2007qx,Voloshin:2003hh}.

\section{\label{sec:NREFT} Soft Photon Decay Spectra}

NREFTs provide a systematic way of incorporating binding effects in the computation of bound state decay amplitudes. One computes the decay amplitudes in electroweak theory. Then a NREFT Hamiltonian is constructed to reproduce the soft photon limit of the electroweak amplitudes when ignoring binding effects. In other words, the effective theory dynamics (ignoring binding effects)  are set equal to the low-energy limit of the electroweak dynamics. The soft-photon limit of the electroweak amplitudes are calculated in Sec.~\ref{sec:soft tree electroweak}. They are used in Secs.~\ref{sec:p-Ps eff theory} and \ref{sec:o-Ps eff theory} to calculate the matching condition used to verify that the effective theory amplitudes (without binding) are indeed equal to the soft photon limit of the electroweak amplitudes.

Once this matching  has been performed, the NREFT Hamiltonian is used to calculate the effective theory amplitudes and subsequently the soft photon spectra. The effective theory amplitudes are calculated using time-ordered perturbation theory and have both long (Coulomb) and short distance (annihilation into a $\nu_\ell \bar{\nu}_\ell$ pair) contributions.

The Coulomb ($H_C$) and Coulomb interaction ($H_\text{int}$) Hamiltonians describe the bound state dynamics of an $e^+e^-$ pair interacting with a quantized electromagnetic field. Following Ref.~\cite{Voloshin:2003hh}, we argue that the dipole approximation of the Coulomb interaction Hamiltonian is valid in the energy range $E_\gamma \ll m$ (Sec.~\ref{sec:dipole}). In the dipole approximation, the Coulomb Hamiltonians are 
\begin{eqnarray}
	H & = & H_\text{C}+H_\text{int}, 
	\label{eq:Hamiltonian}
	\\
	H_\text{C} & = & \frac{\mathbf{p}^2}{m} - \frac{\alpha}{r},
	\\
	H_\text{int} & = & -e\mathbf{x} \cdot \mathbf{E} 
			- \mu \left[ \boldsymbol{\sigma}_{\phi} 
			+ \boldsymbol{\sigma}_{\chi} \right] \cdot \mathbf{B},
	\label{eq:Hint}
\end{eqnarray}
in terms of the center of mass variables $\mathbf{p}=\left(\mathbf{p}_{1}-\mathbf{p}_{2}\right)/2$  and $\mathbf{x}=\mathbf{x}_{1}-\mathbf{x}_{2}$ where the subindices 1, 2 refer to the electron and positron  \cite{Manohar:2003xv}. Here,  $\boldsymbol{\sigma}_{\phi/\chi}$ are the Pauli matrices acting on the electron ($\phi$) and positron ($\chi$) spinors. The electric, $\mathbf{E}$ and magnetic, $\mathbf{B}$, fields are evaluated in the dipole approximation and $H_\text{int}$ can induce both E1 and M1 transitions within the Ps atom.

The Coulomb Hamiltonian $H_C$ is the leading term in the velocity of the electron $v \ll 1$.
The Coulomb interaction Hamiltonian, $H_\text{int}$, is higher order in $v$ and taken as a perturbation. 
The (p/o)-Ps annihilation amplitude is given by the first order $v$ expansion of the electroweak $e^+e^-\to\nu_\ell\bar{\nu}_\ell$ annihilation amplitude calculated in Appendix \hyperref[sec:appC]{C}. While the neutrino energies are of order $\O(m)$, a non-relativistic treatment is still valid since the annihilation into a neutrino-antineutrino pair is a short distance effect -- the neutrinos are not dynamical.

\subsection{\label{sec:soft tree electroweak} Soft Photon Limit of the Tree Level Electroweak Decay Amplitude}

Using the standard Feynman rules, the $\text{Ps}\to\gamma\nu\bar{\nu}$ decay amplitude (Fig.~\ref{fig:gnn diagrams}) is 
\begin{equation}
	\mathcal{M} = -2\sqrt{2} i \GF e m \bar{v} \left( p_2 \right)
		\left(
			\cancel{J} \left( v_\ell- a_\ell \gamma^5 \right)
				\frac{ \cancel{p}_1-\cancel{k}_\gamma+m }{ (p_1-k_\gamma)^2-m^2 } 
					\cancel{\epsilon}_\gamma^{*}
			+ \cancel{\epsilon}_\gamma^* \frac{ \cancel{k}_\gamma-\cancel{p}_2+m }{ (k_\gamma-p_2)^2-m^2 } 
				\cancel{J} \left( v_\ell- a_\ell \gamma^5 \right)
		\right) u\left(p_{1}\right),
	\label{eq:full amp}
\end{equation}
where  $J^{\mu}\left(k_{1},k_{2}\right)=\bar{u}(k_{1})\gamma^{\mu}\left(1-\gamma^{5}\right)v(k_{2})$ is the neutrino current, $p_1$ and $p_2$ are the electron and positron 4-momenta, $k_{1}$ and $k_{2}$ are the neutrino and antineutrino 4-momenta, $k_\gamma$ is the photon 4-momentum and $\epsilon_\gamma$ is the photon polarization.

We choose the Dirac representation for the electron and positron spinors in \eqref{eq:full amp}. In this representation, the electron spinor is
\begin{equation}
	u_{s}(\mathbf{p})
	= \frac{1}{\sqrt{E+m}}
			\left(
				\begin{array}{c}
					E+m 
					\\
					\mathbf{p} \cdot \boldsymbol{\sigma}
				\end{array}
			\right) \phi_{s},
\end{equation}
where  $E=\sqrt{m^2+\mathbf{p}^2}$, $\phi_s$ is the two-component electron spinor and the index $s$ denotes the spin projection \cite{dick2016advanced}. The positron spinors are related to the electron spinors by charge conjugation, 
\begin{equation}
	v_{s}\left(\mathbf{p}\right)
	= \frac{1}{\sqrt{E+m}}
		\left(
			\begin{array}{c}
				\mathbf{p}\cdot\boldsymbol{\sigma}
				\\
				E+m
			\end{array}
		\right)\chi_{s},
\end{equation}
where $\chi_s$ is the two-component spinor of the positron.

Since the Ps binding energy is small, $\O(m\alpha^2)$, the typical momentum of the electron is small  and we neglect it (i.e., $p_{1}=p_{2}=(m,\mathbf{0})$). In the limit $E_\gamma\to0$, the neutrino momenta are back to back ($\mathbf{k}_{1}=-\mathbf{k}_{2}$) and $J^{0}\to0$. Factoring out the $E_\gamma$ dependence and working with $\hat{\mathbf{k}}_\gamma=\mathbf{k}_\gamma/E_\gamma$, equation (\ref{eq:full amp}) becomes 
\begin{equation}
	\mathcal{M} =
		2\sqrt{2} \GF e \chi^{\dagger} 
			\left(
			 	v_\ell \left( \hat{\mathbf{k}}_\gamma \times \boldsymbol{\epsilon}_\gamma \right)
					\cdot\mathbf{J}
				+  a_\ell \left( \boldsymbol{\epsilon}_\gamma \times \mathbf{J} \right)
					\cdot\boldsymbol{\sigma}		
			\right) \phi,
\end{equation}
where we choose $\epsilon_\gamma$ to be real and transverse to $k_\gamma$. Projecting the electron and positron spinors onto the p-Ps ($\chi^{\dagger} \phi \to \sqrt{2}$ and $\chi^{\dagger} \boldsymbol{\sigma} \phi \to\mathbf{0}$)  and o-Ps  ($\chi^{\dagger}\phi\to0$ and $\chi^{\dagger} \boldsymbol{\sigma} \phi \to \sqrt{2} \boldsymbol{\xi}$) states, the low-energy limit of the electroweak amplitudes are
\begin{eqnarray}
	\mathcal{M}_{\text{p}} 
	&=& 4 \GF e v_\ell
	\left(\boldsymbol{\epsilon}_\gamma \times \mathbf{J} \right) \cdot \hat{\mathbf{k}}_\gamma,
	\label{eq:Ap full QFT}
	\\
	\mathcal{M}_{\text{o}} &=& 
		 4 \GF e a_\ell
		\left(\boldsymbol{\epsilon}_\gamma\times\mathbf{J}\right) \cdot \boldsymbol{\xi},
	\label{eq:Ao full QFT}
\end{eqnarray}
where $\boldsymbol{\xi}$ is the o-Ps polarization vector.

\subsection{\label{sec:dipole} Dipole Approximation of the Coulomb Interaction Hamiltonian}

While normally the dipole approximation is applicable for photons with wavelengths much larger than the spatial extent of the Ps atom, $2/m\alpha$ (i.e., $E_\gamma \ll m\alpha$), it has been shown that the dipole approximation of the Coulomb interaction Hamiltonian holds in the enlarged energy region $E_\gamma \ll m$ for the three-body decay $\oggg$ \cite{Voloshin:2003hh,RuizFemenia:2007qx}.  In this energy region, amplitudes where the intermediate states propagate via the Coulomb Green's function, are a series in $\alpha \sqrt{m/E_\gamma}\sim\sqrt{\alpha}$ rather than integer powers of $\alpha$.  The main contributions to the effective field theory amplitudes arise from distances of order $\mathcal{O}(1/\sqrt{mE_\gamma})$, which are much smaller than the Ps radius $\mathcal{O}(1/m\alpha)$ \cite{RuizFemenia:2007qx}.  We argue that the same considerations apply to $\gnn$ decays and that the dipole approximation holds in the extended energy range $E_\gamma \ll m$.

\begin{figure}[h]
    \subfloat[]{\label{fig:p-Ps soft gnn}
         \includegraphics[width=0.3\textwidth]{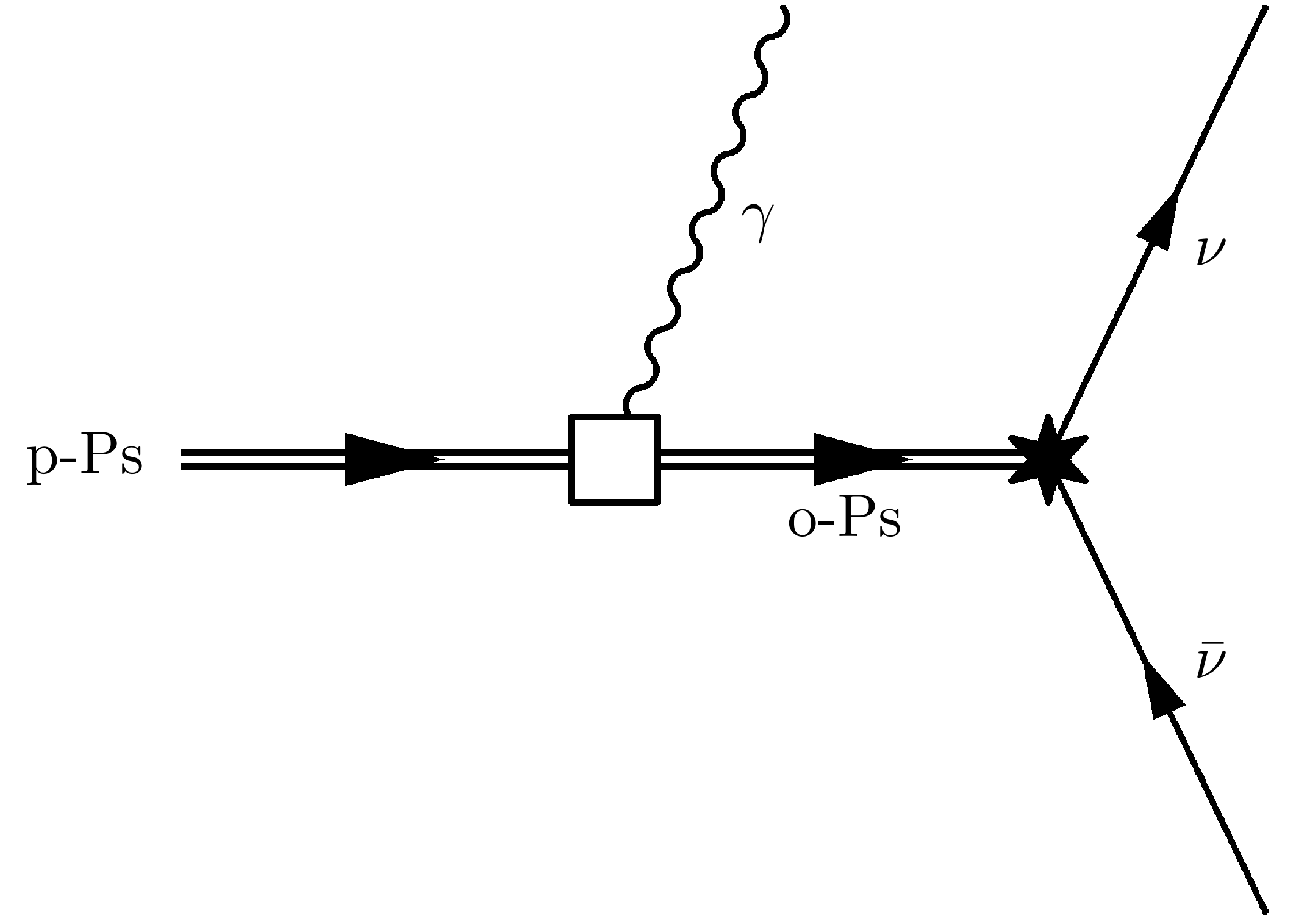}
	}
    \subfloat[]{ \label{fig:o-Ps soft gnn}
        \includegraphics[width=0.3\textwidth]{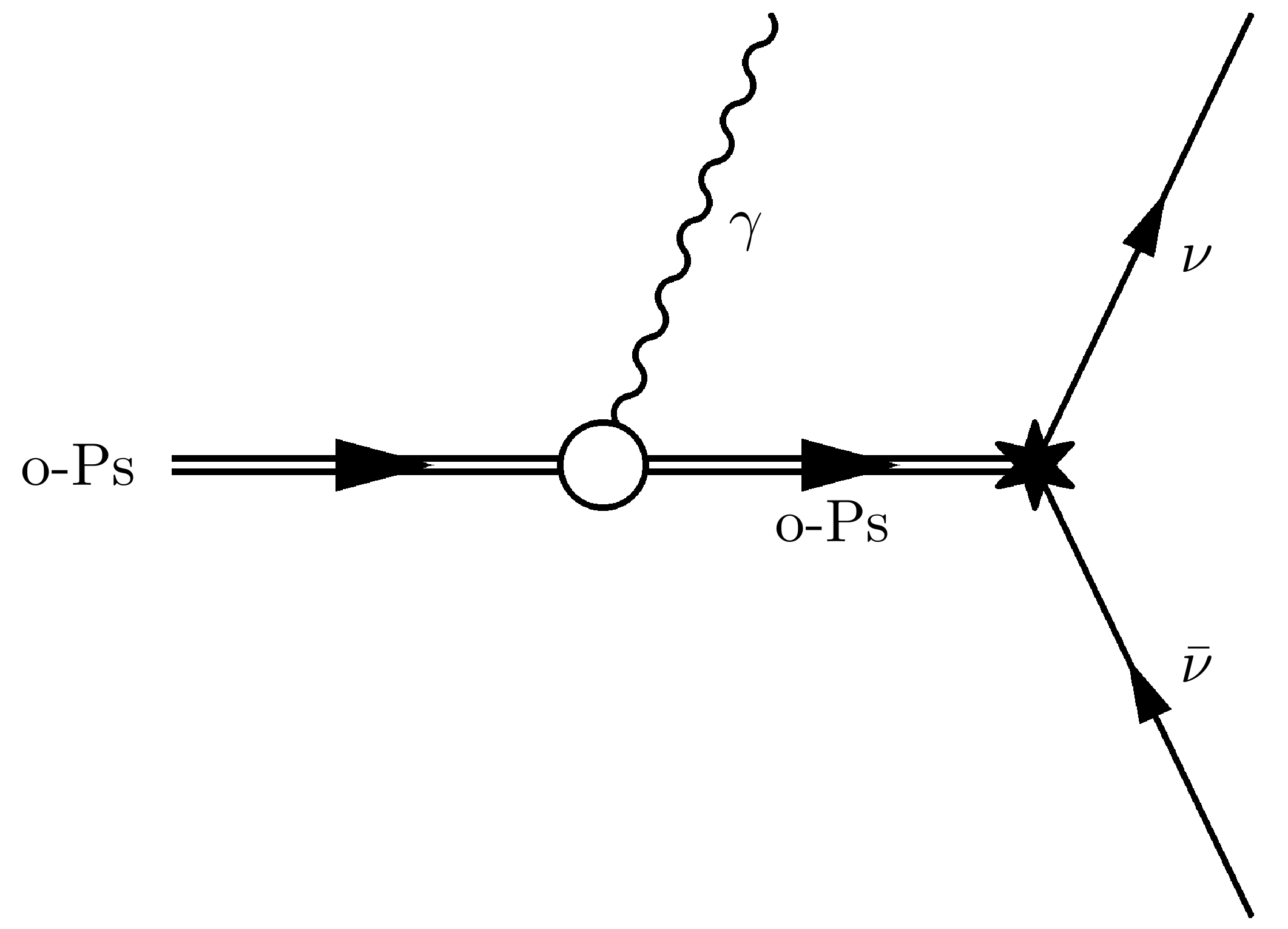}
       }
     \caption{ \label{fig:soft gnn}
     Effective theory graphs for \protect\subref{fig:p-Ps soft gnn} $\pgnn$  
     and \protect\subref{fig:o-Ps soft gnn} $\ognn$. 
     The open square (circle) represents a M1 (E1) transition while the 
     solid star represents the annihilation of o-Ps into 	
     a neutrino-antineutrino pair.
     }
\end{figure}

Initially, the Ps atom is in either the $^1S_0$ or $^{3}S_{1}$ states at energy $E_{0}=-m\alpha^{2}/4$ relative to the threshold.  The p-Ps (o-Ps) atom then emits a soft photon and the $e^{+}e^{-}$ pair propagates non-relativistically in the Coulomb field in a C-odd (C-even) state of energy $E_{0}-E_\gamma$ before annihilating into a neutrino-antineutrino pair (Fig.~\ref{fig:soft gnn}).

The Green's function of the $e^{+}e^{-}$ pair, interacting via a Coulomb field, $G_{C}$, describes the propagation of the pair  between the emission of the soft photon and the annihilation into a neutrino-antineutrino pair. It satisfies the equation
\begin{equation}
	\left(H_{C}+\frac{\kappa^{2}}{m}\right)G_{C}({\bf x},{\bf y};\kappa)=\delta\left({\bf x}-{\bf y}\right)
\end{equation}
and has a factor $\exp\left(-\kappa r\right)$ with
$-\kappa^{2}/m=E_\gamma$. Therefore, the virtual pair propagates over
a distance of $\mathcal{O}(\kappa^{-1})$ \cite{Voloshin:2003hh}.

Since the spin-singlet state cannot annihilate into a
neutrino-antineutrino pair \cite{Czarnecki:1999mt}, the virtual C-odd
(C-even) state of Fig.~\protect\subref*{fig:p-Ps soft gnn}
(\protect\subref*{fig:o-Ps soft gnn}) must be a triplet state of
orbital angular momentum $L=2n$ ($L=2n+1$) for $n$ a non-negative
integer.  The amplitude for annihilation contains $L$ derivatives of
the wave function at the origin and is proportional to
$(\kappa / m)^L$.  Since the $\exp(-\kappa r)$ dependence of the
Green's function constrains the product $E_\gamma r$ to order one, the
contributions of the intermediate states of Fig.~\ref{fig:soft gnn} to
the amplitude are proportional to $(E_\gamma / m)^L$.  

Therefore, only the intermediate states with the lowest $n$ (i.e.,
$n=0$) need to be considered for $E_\gamma \ll m$
\cite{Voloshin:2003hh}.  The intermediate state of
Fig.~\protect\subref*{fig:p-Ps soft gnn} is the o-Ps ground state,
$1^3S_1$, while the intermediate states of Fig.~\ref{fig:o-Ps soft
  gnn} are the $L=1$ o-Ps excited states, $n^3P_{0,1,2}$. These states
are reached from the initial p-Ps and o-Ps ground states by M1 and E1
transitions respectively. Thus, the dipole approximation is valid in
the energy region $E_\gamma \ll m$.

\subsection{\label{sec:p-Ps eff theory} Soft Photon Spectrum for $\boldsymbol{\pgnn}$}

As noted in Sec.~\ref{sec:dipole}, p-Ps cannot decay into a
$\nu_\ell \bar{\nu}_\ell$ pair; therefore, $\pgnn$ decay proceeds
solely through an M1 transition.  The M1 interaction flips the spin of
either the electron or positron and takes the initial p-Ps state,
$1^1S_0$, to an intermediate o-Ps state. Within the dipole
approximation, the only allowed intermediate state is the o-Ps ground
state, $1^3S_1$.

In time-ordered perturbation theory, the effective theory amplitude
for $\pgnn$, Fig.~\protect\subref*{fig:p-Ps soft gnn}, is
\begin{eqnarray}
	\M_\text{p}^\text{eff} & = &  \sum_n 
		\frac{
			i \langle 0 \vert \hat{A}^{\left(\nu_\ell \bar{\nu}_\ell \right)}_\text{s} \vert n \rangle \langle n \vert i \mu 
				\left( \boldsymbol{\sigma}_{\phi} + \boldsymbol{\sigma}_{\chi} \right) 
					\cdot \mathbf{B} \vert \text{p-Ps} \rangle 
		}{ 
			E_\text{p} - E_n - E_\gamma
		}
	\nonumber \\
 	& = &\sum_{m_s} 
		\frac{
			-i \langle 0 \vert \hat{A}^{\left(\nu_\ell \bar{\nu}_\ell\right)}_\text{s} \vert 1^{3}S_{1};m_s \rangle 
				\langle1^{3}S_{1};m_s \vert i \mu 
					\left( \boldsymbol{\sigma}_{\phi}+\boldsymbol{\sigma}_{\chi}\right)
						\cdot	\mathbf{B} \vert \text{p-Ps} \rangle
		}{
			\Delta E_{\text{hfs}} + E_\gamma
		},
	\label{eq:Ap NREFT initial}
\end{eqnarray}
where $\Delta E_\text{{hfs}}=E_\text{o}-E_\text{p}$ is the hyperfine splitting energy difference, and, $E_\text{p}$ and $E_\text{o}$ are the p-Ps and o-Ps ground state energies. Here, $\hat{A}^{(\nu\bar{\nu})}_\text{s}$, is the s-wave $\onn$ annihilation operator (derived in Appendix \hyperref[sec:appC]{C}),
\begin{equation} 
	\hat{A}^{(\nu_\ell \bar{\nu}_\ell)}_\text{s}
		= 2 \sqrt{2} i \GF m  v_\ell \left( \mathbf{J} \cdot \boldsymbol{\sigma} \right).
	\label{eq:p-Ps ann op}
\end{equation}  
To simplify the effective theory amplitude, we begin by evaluating the annihilation and magnetic matrix elements in the numerator. Projecting the electron and positron spinors onto the spin triplet state ($\chi^{\dagger}\boldsymbol{\sigma}\phi \to \sqrt{2} \boldsymbol{\xi}$), the annihilation matrix element becomes
\begin{eqnarray}
	\langle 0 \vert  \hat{A}^{(\nu_\ell \bar{\nu}_\ell)}_\text{s} \vert 1^3S_1;m_s \rangle 
	& = & 2\sqrt{2} i \GF m  v_\ell \mathbf{J} \cdot \left(\chi^{\dagger}\boldsymbol{\sigma}\phi\right) \psi_0(0)
	\nonumber \\
	& = & 4 i \GF m  v_\ell \mathbf{J} \cdot \boldsymbol{\xi} \psi_0(0),
\end{eqnarray}
where $\psi_0$ is the Ps ground state wavefunction. The magnetic matrix element is 
\begin{eqnarray}
	\langle 1^3S_1;m_s \vert i\mu \left( \boldsymbol{\sigma}_{\phi}+\boldsymbol{\sigma}_{\chi} \right)
	\cdot \mathbf{B} \vert \text{p-Ps} \rangle 
	& = & \frac{e}{\sqrt{2}m}E_\gamma
			\left(\hat{\mathbf{k}}_\gamma \times \boldsymbol{\epsilon}_\gamma \right)
				\cdot \left(\chi^{\dagger}\boldsymbol{\sigma}\phi\right)^{*}
	\nonumber \\
	& = & \frac{e}{\sqrt{2}m}E_\gamma
			\left(\hat{\mathbf{k}}_\gamma \times \boldsymbol{\epsilon}_\gamma \right)
				\cdot \sqrt{2}\boldsymbol{\xi}^{*}.
\end{eqnarray}
Summed over the polarizations of the intermediate o-Ps states in \eqref{eq:Ap NREFT initial}, 
\begin{equation}
	\sum_{\boldsymbol{\xi}} \boldsymbol{\xi}^i \boldsymbol{\xi}^{i*} =  \delta^{ij},
\end{equation}
the effective theory amplitude becomes
\begin{eqnarray}
	\M_\text{p}^\text{eff}
	& = & 4 \GF e v_\ell \psi_0(0) 
			\left(\boldsymbol{\epsilon}_\gamma \times \mathbf{J} \right) \cdot \hat{\mathbf{k}}_\gamma
			 	\; \mathcal{A}_\text{m}(E_\gamma),
	\label{eq:Ap NREFT}
\end{eqnarray}
where $\psi_0$ is the ground state Ps wave function. The magnetic
amplitude, $\mathcal{A}_\text{m}$, contains all of the dependence on
soft photon energy in the effective theory amplitude,
\begin{equation}
	\mathcal{A}_\text{m}(E_\gamma)
	= \frac{E_\gamma}{\Delta
          E_{\text{hfs}}+E_\gamma}=\frac{x_\gamma}{\epsilon+x_\gamma},
        \quad \epsilon\equiv \frac{\Delta E_{\text{hfs}}}{m}.
\end{equation}
To ensure that the effective theory amplitude \eqref{eq:Ap NREFT} is
consistent with electroweak theory, we consider 
$x_\gamma \gg \epsilon$ and neglect the hyperfine
energy difference in the energy denominator of \eqref{eq:Ap NREFT
  initial} (i.e., $\mathcal{A}_{m}=1$). The effective theory
amplitude, ignoring binding effects, is therefore
\begin{eqnarray}
	 \M_\text{p}^\text{eff} \to 4 \GF e v_\ell \psi_0(0) 
	 \left(\boldsymbol{\epsilon}_\gamma \times \mathbf{J} \right) \cdot \hat{\mathbf{k}}_\gamma.
	\label{eq:Ap no binding}
\end{eqnarray}
Since \eqref{eq:Ap no binding} is equal to the soft photon limit of the tree
level electroweak amplitude \eqref{eq:Ap full QFT}, 
the M1 transition and annihilation operator
\eqref{eq:p-Ps ann op} fully account for the emitted soft photon and $\nu_\ell \bar{\nu}_\ell$
annihilation in $\pgnn$ decays.

Assured that the effective theory amplitude \eqref{eq:Ap NREFT} is
consistent with the full electroweak theory, we use it to calculate
the low-energy photon spectrum. We need both the three body
phase space in the $x_\gamma\to0$ limit and the spin averaged
amplitude squared. In the $x_\gamma\to 0$ limit, the three body phase
space is
\begin{equation}
	\left[ \frac{1}{128\pi^{3}} \dd x_{1} \dd x_\gamma \right]_{x_\gamma \to 0}
		\approx \frac{1}{128\pi^{3}}\frac{x_\gamma}{2} \dd \cos\theta \dd x_\gamma,
	\label{eq:low energy PS}
\end{equation}
where $\theta$ is the angle between the neutrino and photon. The spin averaged square of the amplitude is
\begin{eqnarray}
	\sum_{\boldsymbol{\epsilon}_\gamma} \left| \M^\text{eff}_\text{p} \right|^2
	& = & \sum_{\boldsymbol{\epsilon}_\gamma}
		\left| 4 \GF e v_\ell
			\psi_0(0)	\mathcal{A}_\text{m} (E_\gamma)
			\left(\boldsymbol{\epsilon}_\gamma 
                       \times \mathbf{J} \right) \cdot \hat{\mathbf{k}}_\gamma \right|^2
	\nonumber \\
 	& _{x_\gamma\to0}\to & 128 \GF^2 v^2_\ell \alpha^4 m^5
		\left| \mathcal{A}_\text{m} (E_\gamma) \right|^2 \left( 1+\cos^2\theta \right),
	\label{eq:Ap NREFT sqrd}
\end{eqnarray}
where
$ \sum_{\boldsymbol{\epsilon}_\gamma} |
\left(\boldsymbol{\epsilon}_\gamma \times \mathbf{J} \right) \cdot
\hat{\mathbf{k}}_\gamma|^2 = 16 E_1^2 (1+(\hat{\mathbf{k}}_\gamma
\cdot \hat{\mathbf{k}}_1 )^2 )$,
$\hat{\mathbf{k}}_\gamma \cdot \hat{\mathbf{k}}_1 = \cos\theta$ and
$E_1 \to m$.  Here, $\hat{\mathbf{k}}_1$ and $\hat{\mathbf{k}}_\gamma$
are the unit 3-momentum vectors of the neutrino and photon.

The effective theory photon spectrum is obtained by multiplying
\eqref{eq:Ap NREFT sqrd} by \eqref{eq:low energy PS} and integrating
over $\dd \cos\theta$ where the allowed integration range is
$-1\leq\cos\theta\leq1$
\begin{eqnarray}
	\left( \frac{1}{\Gamma_\text{p}} \frac{\dd\Gamma_\text{p}}{\dd x_\gamma} \right)^\text{eff}
	& = & \frac{ 9 \pi^3 }{ 2 m^5 \alpha^4 \GF^2 v_\ell^2 }
	\int_{-1}^{1} \dd\cos\theta \frac{1}{128 \pi^3} \frac{x_\gamma}{2}
		\sum_{\boldsymbol{\epsilon}_\gamma} \left| \M_\text{p}^\text{eff} \right|^2
	\nonumber \\
 	& = &6 x_\gamma \left| \mathcal{A}_\text{m} (E_\gamma) \right|^2.
	\label{eq:fp Low}
\end{eqnarray}
The  spectrum is proportional to the square of the
magnetic amplitude, $\mathcal{A}_\text{m}$. The magnetic amplitude has
simple asymptotic behavior; it is linear in $x_\gamma$ for
$x_\gamma \ll \epsilon$ and approximately constant for
$x_\gamma \gg \epsilon$
\begin{equation}
	\mathcal{A}_\text{m} \approx
		\begin{cases}
			x_\gamma / \epsilon & \text{for } x_\gamma \ll \epsilon 
			\\
			1 & \text{for } x_\gamma \gg \epsilon.
		\end{cases}
\end{equation}
Therefore, the effective theory spectrum \eqref{eq:fp Low} is cubic in
$x_\gamma$ in the low-energy limit,
$x_\gamma \ll \epsilon$, as required by Low's theorem.
Above the hyperfine splitting, $x_\gamma\gg\epsilon$, the
spectrum shifts from being cubic in the photon energy to linear.

\begin{figure}[h]
	\centering
	\subfloat[]{	\label{fig:pRatio1}
			\includegraphics[width=.5\textwidth]{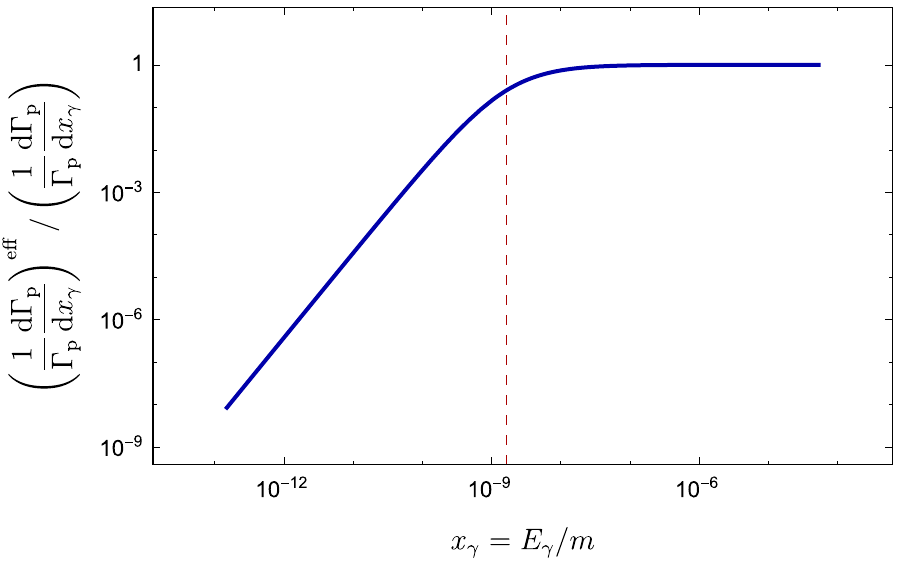}
			}
	\subfloat[]{\label{fig:pRatio2}
		\includegraphics[width=.5\textwidth]{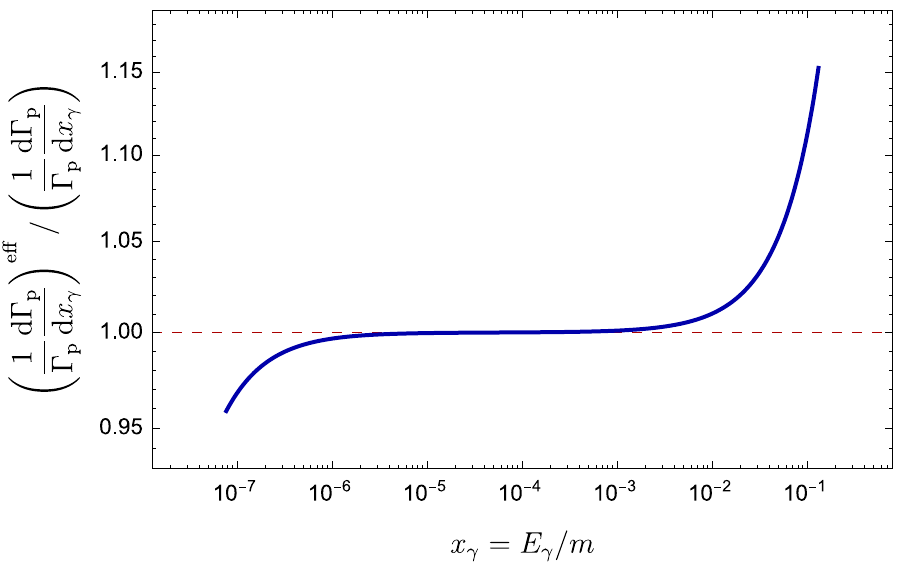}
		}
	\caption{\label{fig:pRatio}
		Log-log plot of the ratio of the effective theory amplitude to the tree 
		level electroweak amplitude for $\text{p-Ps}\to\gamma\nu\bar{\nu}$ 
		decays in \protect\subref{fig:pRatio1} the low-energy limit 
		$\alpha^6 < x_\gamma < \alpha^2$ and \protect\subref{fig:pRatio2}
		the high-energy limit $\alpha^3/5 < x_\gamma < 1$.
		The vertical line in \protect\subref{fig:pRatio1} indicates the 
		hyperfine splitting energy ($x_\gamma =\epsilon = 7\alpha^4/12$)
		while the horizontal line in \protect\subref{fig:pRatio2} is placed at 1 
	to indicate the region where the effective and full theory 
		spectra are equal. 
		}
\end{figure}

The ratio of the $\pgnn$ effective theory to the tree level
electroweak spectrum is plotted in Fig.~\ref{fig:pRatio}. In the
intermediate energy region
($\epsilon \ll x_\gamma \ll 1$), the ratio plateaus near
1 (Fig.~\ref{fig:pRatio}) indicating that the effective theory and
tree level electroweak spectrum \eqref{eq:p-Ps spectrum} are
approximately equal (the two spectra intersect at
$x_\gamma \approx 5.75 \times 10^{-5}$).  For high-energy photons
$x_\gamma \lesssim 1$, the ratio spikes revealing that the effective
theory spectrum differs significantly from the tree level electroweak
spectrum and is no longer accurate
(Fig.~\protect\subref*{fig:pRatio2}).  Below the hyperfine energy
splitting, the ratio in the log-log plot is linear with a slope of $2$
since the effective theory spectrum is cubic in $x_\gamma$ while the
tree level electroweak spectrum is linear
(Fig.~\protect\subref*{fig:pRatio1}).

\subsection{\label{sec:o-Ps eff theory} Soft Photon Spectrum for $\boldsymbol{\text{o-Ps}\to\gamma\nu\bar{\nu}}$}
In $\ognn$ decays, the E1 transition takes the initial o-Ps ground
state, $1^3S_1$, to the excited o-Ps states $n^3P_{0,1,2}$ ($n\neq1$),
which then decay into a $\nu_\ell \bar{\nu}_\ell$ pair.  The M1
transition takes the initial o-Ps state, $1^3S_1$, to the p-Ps ground
state, $1^1S_0$, which cannot decay into a $\nu_\ell \bar{\nu}_\ell$
pair and therefore does not need to be considered.

The effective theory $\ognn$ decay amplitude, Fig.~\protect\subref*{fig:o-Ps soft gnn}, is given by
\begin{eqnarray}
	\M_\text{o}^\text{eff} 
	& = & \sum_{n} 
			\frac{
				i \langle 0 \vert \hat{A}^{\left(\nu_\ell \bar{\nu}_\ell\right)}_\text{p} \vert n \rangle 
					\langle n \vert ie \mathbf{x} \cdot \mathbf{E} \vert \text{o-Ps} \rangle
	}{
		E_{o}-E_{n}-E_\gamma
	}
\nonumber \\
 	& = & - 2\sqrt{2} i \GF e a_\ell E_\gamma \sum_{n} 
		\frac{ 
			\langle 0 \vert \left(\mathbf{J} \times \boldsymbol{\sigma} \right) 
				\cdot \mathbf{p} \vert n \rangle \langle n \vert \mathbf{x} 
					\cdot \boldsymbol{\epsilon}_\gamma \vert \text{o-Ps} \rangle
		}{
			E_\text{o}-E_{n}-E_\gamma
		},
	\label{eq:o-Ps NREFT amp}
\end{eqnarray}
where $\hat{A}^{\left(\nu_\ell \bar{\nu}_\ell \right)}_\text{p} $ is the p-wave $\onn$ annihilation operator (derived in Appendix \hyperref[sec:appC]{C}),
\begin{equation}
	\hat{A}^{\left(\nu_\ell \bar{\nu}_\ell\right)}_\text{p} 
		= - 2 \sqrt{2} \GF a_\ell \left( \mathbf{J}\times\boldsymbol{\sigma} \right) \cdot \mathbf{p}.
	\label{eq:o-Ps ann op}
\end{equation}
As in the calculation of the effective theory $\pgnn$ amplitude
(Sec.~\ref{sec:p-Ps eff theory}), we now demonstrate that the effective theory 
amplitude (without binding) is equal to the soft photon limit of the electroweak amplitude. 
To calculate the effective theory amplitude, ignoring binding effects, we take
$E_\gamma \gg m\alpha^2$ and ignore $ E_\text{o}-E_{n}$ in the energy
denominator of \eqref{eq:o-Ps NREFT amp} which yields
\begin{eqnarray}
	\M_\text{o}^\text{eff}
	& \to & 
	2 \sqrt{2} i \GF a_\ell e \sum_{n} 
		\langle 0 \vert \left(\mathbf{J} \times \boldsymbol{\sigma} \right) 
			\cdot \mathbf{p} \vert n \rangle \langle n \vert \mathbf{x} 
				\cdot \boldsymbol{\epsilon}_\gamma \vert \text{o-Ps} \rangle
	\nonumber \\
	& = &  2 \sqrt{2} i \GF a_\ell e  
		\langle 0 \vert \left(\mathbf{J} \times \boldsymbol{\sigma} \right) 
			\cdot \mathbf{p}  \mathbf{x} 
				\cdot \boldsymbol{\epsilon}_\gamma \vert \text{o-Ps} \rangle.
	\label{eq:Ao no binding 1}
\end{eqnarray}
The tensor operator $\mathbf{p}^i \mathbf{x}^j$ can be decomposed into irreducible spherical tensor operators
\begin{eqnarray}
	\mathbf{p}^i \mathbf{x}^j 
	&=& \frac {\delta^{ij}}{3} \mathbf{p}\cdot\mathbf{x} 
		+ \frac{\mathbf{p}^i \mathbf{x}^j - \mathbf{p}^j \mathbf{x}^i}{2}
			+ \frac{1}{2} 
				\left( 
					\mathbf{p}^i \mathbf{x}^j 
						+ \mathbf{p}^j \mathbf{x}^i 
							- \frac{2}{3} \delta^{ij} \mathbf{p} \cdot \mathbf{x} 
				\right).
	\label{eq:irr tensor decomp}
\end{eqnarray}
Since the initial o-Ps state is an s-wave, only the operator with zero
angular momentum (first term of \eqref{eq:irr tensor decomp}) gives a
non-zero matrix element.  Additionally, we may take the operator
$\mathbf{p}$ to act only on $\mathbf{x}$ because
$\mathbf{x}\cdot\boldsymbol{\nabla}\psi_0$ vanishes at the
origin. With these considerations, the effective theory amplitude
(ignoring binding effects) \eqref{eq:Ao no binding 1} simplifies to
\begin{eqnarray}
	\M_\text{o}^\text{eff} \to 4\GF e a_\ell 
		\left(\mathbf{J} \times \boldsymbol{\epsilon}_\gamma \right) \cdot \boldsymbol{\xi} \; \psi_0(0).
	\label{eq:Ao no binding 2}
\end{eqnarray}
Since this is equal to the soft photon limit of the tree level
electroweak amplitude \eqref{eq:Ao full QFT}, 
the E1 transition and annihilation operator (\ref{eq:o-Ps ann op}) fully
account for the emitted soft photon and 
$\nu_\ell \bar{\nu}_\ell$ annihilation in $\ognn$
decays. Thus, equation \eqref{eq:o-Ps NREFT amp} is the complete
effective theory amplitude.

We now return to the general case, without any assumptions about
photon energies. Expanding the inner products of the effective theory
amplitude \eqref{eq:o-Ps NREFT amp}, we find
\begin{eqnarray}
\label{eq:o-Ps NREFT amp 2}
	\M_\text{o}^\text{eff} 
 	&=&
	4 \GF e a_\ell E_\gamma \left(\mathbf{J} \times \boldsymbol{\xi} \right)^i \boldsymbol{\epsilon}_\gamma^j
	\int \dd^3 x ~ \dd^3 y ~ \delta^{(3)}(\mathbf{x}) ~
	 \partial^i_x 
	\left(
		\sum_{n}
		\frac{ \langle \boldsymbol{x} \vert n \rangle \langle n \vert \boldsymbol{y} \rangle}{E_{n}+\kappa^2/m} 
	\right)
	y^j \psi_0(y)
	\nonumber \\
	&=&
	4 \GF e a_\ell E_\gamma \left(\mathbf{J} \times \boldsymbol{\xi} \right)^i \boldsymbol{\epsilon}_\gamma^j
	\int \dd^3 y ~ 
	 \left[\partial^i_x G_C(\mathbf{x},\mathbf{y},\kappa)\right]_{\mathbf{x}=\mathbf{0}} y^j \psi_0(y)		
\end{eqnarray}
where $-\kappa^2/m = E_\text{o}-E_\gamma$ and $G_C(\mathbf{x},\mathbf{y},\kappa)$ is the Coulomb Green's function. The derivative selects the $l=1$ partial wave of the Green's function \cite{RuizFemenia:2007qx}
\begin{equation}
\label{eq:dG}
	\left[\partial^i_x G_C(\mathbf{x},\mathbf{y},\kappa)\right]_{\mathbf{x}=\mathbf{0}} = 3y^i G_1(0,y,\kappa).
\end{equation}
where the partial wave decomposition of the Coulomb Green's function
can be found in Appendix C of Ref.~\cite{Manohar:2003xv}. Substituting
\eqref{eq:dG} into \eqref{eq:o-Ps NREFT amp 2} and preforming the
angular integrations yields the effective theory amplitude
\begin{eqnarray}
\label{eq:Ao NREFT}	
	\M_\text{o}^\text{eff} 
	&=&
	4\GF e a_\ell \left(\mathbf{J} \times \boldsymbol{\xi} \right) \cdot \boldsymbol{\epsilon}_\gamma \psi_0(0)
	\mathcal{A}_e(E_\gamma).	
\end{eqnarray}
 Here, the electric amplitude, $\mathcal{A}_\text{e}$, is determined to be
\begin{eqnarray}
	\mathcal{A}_\text{e} \left(E_\gamma\right)
	& = & \frac{4 \pi E_\gamma}{\psi_0(0)} \int_0^\infty \dd y \, y^4 G_{C,1} \left(0,y;\kappa \right) \psi_0(y)
	\nonumber \\
 	& = & \frac{
			\left(1-\nu\right) \left(3+5\nu\right) 
		}{
			3\left(1+\nu\right)^{2}
		}
		+ \frac{
			8\nu^{2} \left(1-\nu\right)
		}{
			3\left(2-\nu\right)\left(1+\nu\right)^{3}
		}
			\;_{2}F_{1} \left(1,2-\nu; 3-\nu; \frac{\nu-1}{\nu+1}\right),
	\label{eq:Ae}
\end{eqnarray}
where 
$\nu = \frac{\alpha}{\sqrt{4x_\gamma+\alpha^2}}$.
In the first line of
\eqref{eq:Ae} we use 
the integral representation of the electric
amplitude from  Ref.~\cite{Voloshin:2003hh}.  The hypergeometric
function $_2F_{1}$ simplifies to the
so-called Hurwitz-Lerch $\Phi$ function \cite{NISTHandbook}, 
\begin{eqnarray}
\frac{1}{2-\nu}\,{_{2}F_{1}}\left(1,2-\nu; 3-\nu;
  \frac{\nu-1}{\nu+1}\right) &=& 
\frac{1}{2-\nu}\Phi\(\frac{\nu-1}{\nu+1},1,2-\nu\)
=\sum_{n=0}^\infty \frac{1}{2-\nu + n} \( \frac{\nu-1}{\nu+1}\)^n,
\\
	\mathcal{A}_\text{e} \left(E_\gamma\right)
 	& = & \frac{ 1-\nu}{ 3\( 1+\nu \)^2}
\[3+5\nu  +  \frac{8 \nu^2}{ 1+\nu } \sum_{n=0}^\infty \frac{1}{2-\nu + n} \( \frac{\nu-1}{\nu+1}\)^{n}\].
	\label{eq:Ae2}
\end{eqnarray}
At high energies, equivalent to $x_\gamma \gg \alpha^2$ and
$\nu\simeq \frac{\alpha}{2\sqrt{x_\gamma}}\to 0$, this amplitude can
be expanded as a series in $\alpha/\sqrt{x_\gamma}$,
\begin{equation}
	\mathcal{A}_\text{e} = 1 - \frac{2\alpha}{3\sqrt{x_\gamma}} + \frac{
          \left(2 - 2\ln 2\right) \alpha^2}{3x_\gamma} + \dots, \quad  (x_\gamma \gg \alpha^2) .
	\label{eq:x>>alpha2}
\end{equation}
For  $x_\gamma \gg \alpha^2$, the electric amplitude is thus
approximately $1$.  In this region the binding effects are relatively
unimportant. Indeed, the expression
\eqref{eq:Ao NREFT} agrees with the amplitude obtained when binding
effects are ignored, eq.~\eqref{eq:Ao no binding 2}, when we take $\mathcal{A}_e \to 1$.

On the other hand, in the extreme soft photon limit $x_\gamma \ll
\alpha^2$, equivalent to $\nu\simeq 1-\frac{2x_\gamma}{\alpha^2}\to 1^-$, the electric amplitude can be
expanded as a series in $x_\gamma/\alpha^2$. The leading behaviour is
\begin{equation}
	\mathcal{A}_\text{e} = \frac{2x_\gamma}{\alpha^2}
+ \dots \quad  (x_\gamma \ll \alpha^2) .
	\label{eq:x<<alpha2}
\end{equation}
The leading term in the soft photon limit is linear in $x_\gamma$ with
a slope of $2/\alpha^2$. 

To summarize, the electric amplitude is linear in the photon energy below the binding energy and approximately constant above it.
The expansions \eqref{eq:x<<alpha2} and \eqref{eq:x>>alpha2} will be important when determining the behaviour of the photon spectrum in the limits $x_\gamma\ll\alpha^2$ and $x_\gamma\gg\alpha^2$.

It is instructive to look for a simpler way to derive the
leading low-energy term \eqref{eq:x<<alpha2}. In the soft photon
limit, the wavelength is large and the electric field of the wave is
approximately constant. This is similar to the situation in the Stark
effect. 
Since the first order correction to the ground state
energy for the Stark effect vanishes
($E^{(1)} \propto \langle \psi_0\vert \mathbf{x}\cdot
\boldsymbol{\epsilon}_\gamma \vert \psi_0 \rangle = 0$),
one evaluates the second order correction to the ground state energy
\begin{equation}
	E^{(2)} =
		\sum_{n\neq0} \frac{
					\langle\psi_0\vert H^\prime \vert n\rangle\langle n\vert H^\prime \vert\psi_0\rangle
				}{E_0-E_n},
	\label{eq:E2stark}
\end{equation}
where
$H^\prime \propto \mathbf{x} \cdot \boldsymbol{\epsilon}_\gamma = r
\cos\theta$.
The form of \eqref{eq:E2stark} is similar to the low-energy limit of
the effective theory amplitude where $E_\gamma=0$ in the energy
denominator of \eqref{eq:o-Ps NREFT amp}
\begin{eqnarray}
	\M_\text{o}^\text{eff} 
 	& = & - 2\sqrt{2} i \GF e a_\ell E_\gamma \sum_{n} 
		\frac{ 
			\langle 0 \vert \left(\mathbf{J} \times \boldsymbol{\sigma} \right) 
				\cdot \mathbf{p} \vert n \rangle \langle n \vert \mathbf{x} 
					\cdot \boldsymbol{\epsilon}_\gamma \vert \text{o-Ps} \rangle
		}{
			E_\text{o}-E_{n}
		}.
	\label{eq:D-L}
\end{eqnarray}
Since equation \eqref{eq:E2stark} can be summed exactly using the
method of Dalgarno and Lewis \cite{DalgarnoLewis,schwartz:1959aa}, we
can exploit the similarity between equations \eqref{eq:E2stark} and
\eqref{eq:D-L} to evaluate the effective theory amplitude in the soft
photon limit.

Equations \eqref{eq:E2stark} and \eqref{eq:D-L} can be summed exactly by finding a function $F$ that satisfies
\begin{equation}
	[F,H_0] \psi_0(\mathbf{x}) = \mathbf{x}\cdot \boldsymbol{\epsilon}_\gamma \psi_0(\mathbf{x}).
\end{equation}
For the unperturbed positronium Hamiltonian, $H_0$, the function $F$ is given by
\begin{equation}
	F = -\frac{m}{2} \mathbf{x}\cdot \boldsymbol{\epsilon}_\gamma \left( a^2 +\frac{ar}{2} \right).
\end{equation}
With $F$ in hand, we evaluate equation \eqref{eq:D-L} 
\begin{eqnarray}
	\M_\text{o}^\text{eff} 
 	& = & - 2\sqrt{2} i \GF e a_\ell E_\gamma 
			\langle 0 \vert \left(\mathbf{J} \times \boldsymbol{\sigma} \right) 
				\cdot \mathbf{p} F \vert \text{o-Ps} \rangle
	\nonumber \\
	& = & - 4 \GF e a_\ell E_\gamma 
			 \left(\mathbf{J} \times \boldsymbol{\xi} \right) 
				\cdot \int \dd^3 x \; \delta^{(3)}(\mathbf{x}) \boldsymbol{\nabla} (F\psi_0(\mathbf{x}))
	\nonumber \\
	& = &  4 \GF e a_\ell \psi_0(0)
			 \left(\boldsymbol{\epsilon}_\gamma \times \mathbf{J} \right) 
				\cdot  \boldsymbol{\xi}  \frac{2x_\gamma}{\alpha^2}. 
\end{eqnarray}
Thus, in the limit $x_\gamma \ll \alpha^2$ the electric amplitude is $\mathcal{A}_e \approx 2x_\gamma / \alpha^2$ which is equal to the first order term of the expansion \eqref{eq:x<<alpha2}. 

Similarly, the Stark effect can be related to the soft photon limit of
the E1 portion of the $\oggg$ decay amplitude. The annihilation
operator that contributes to the E1 portion of the $\oggg$ decay
amplitude is of the same form as the o-Ps p-wave
$\nu_\ell\bar{\nu}_\ell$ annihilation operator and contains a $\mathbf{p}$ derivitive.
 A calculation, using the summation
technique above, reveals that in the soft photon limit,
$\mathcal{A}_e\approx 2x_\gamma/\alpha^2$. This agrees with the soft
photon limit of the electric amplitude derived in
\cite{Manohar:2003xv,RuizFemenia:2007qx} by expansion of the
p-wave Green's function.

With this understanding of the electric amplitude, we proceed to
the photon spectrum. Both the spin averaged square of the
amplitude \eqref{eq:Ao NREFT} and the three body phase space in the
$x\to0$ limit are needed. Squaring \eqref{eq:Ao NREFT}, summing over
the photon polarizations and averaging over the initial o-Ps
polarizations, yields
\begin{eqnarray}
	\frac{1}{3} \sum_{\boldsymbol{\xi} \boldsymbol{\epsilon}_\gamma} \left| \M_\text{o}^\text{eff}  \right|^2
		&=& \frac{1}{3} \sum_{\boldsymbol{\xi} \boldsymbol{\epsilon}_\gamma}
				\left| 4\GF e a_\ell \psi_0(0)
					\mathcal{A}_\text{e}(E_\gamma) \left( \boldsymbol{\epsilon}_\gamma 
						\times \mathbf{J} \right) \cdot \boldsymbol{\xi} \right|^2
	\nonumber \\
		&_{x_\gamma\to0}\to& 128 \GF^2 a_\ell^2 \alpha^4  m^5  
			\left| \mathcal{A}_\text{e}(E_\gamma) \right|^2 
				\left( 1-\frac{1}{3} \cos^2\theta \right),
\end{eqnarray}
where 
$	
	\sum_{\boldsymbol{\xi}\boldsymbol{\epsilon}_\gamma} 
	|
		\left( \boldsymbol{\epsilon}_\gamma \times \mathbf{J} \right)
						\cdot \boldsymbol{\xi} 
	|^2 /3
	=(16E_{1}^{2}) (1-\frac{1}{3} ( \hat{\mathbf{k}}_\gamma \cdot \hat{\mathbf{k}}_1 )^2 )
$,
$\hat{\mathbf{k}}_\gamma\cdot\hat{\mathbf{k}}_{1}=\cos\theta$ and $E_1\to m$. Multiplying by the three body phase space in the limit $x_\gamma\to0$ and integrating over $\cos\theta$ yields the effective theory spectrum
\begin{eqnarray}
	\left( \frac{1}{\Gamma_\text{o}} \frac{\dd \Gamma_\text{o}}{\dd x_\gamma} \right)^\text{eff}
	& = & \frac{27\pi^3} {8 \GF^2 m^5 \alpha^4 a_\ell^2}
			\int_{-1}^{1} \dd\cos\theta \frac{1}{128 \pi^3} \frac{x_\gamma}{2} 
				\frac{1}{3} \sum_{\boldsymbol{\xi} \boldsymbol{\epsilon}_\gamma}
					\left| \M_\text{o}^\text{eff} (E_\gamma) \right|^2
	\nonumber \\
 	& = & 3 x_\gamma \left| \mathcal{A}_\text{e}(E_\gamma) \right|^2.
	\label{eq:fo Low}
\end{eqnarray}
The effective theory spectrum is proportional to the square of the electric amplitude and thus shares the same transitional behaviour at $x_\gamma=\alpha^2$. Substituting the leading term from equations \eqref{eq:x<<alpha2} and \eqref{eq:x>>alpha2} into \eqref{eq:fo Low} we obtain the approximate form of the spectrum in the limits $x_\gamma\ll\alpha^2$ and $x_\gamma\gg\alpha^2$ 
\begin{equation}
	\left( \frac{1}{\Gamma_\text{o}} \frac{\dd \Gamma_\text{o}}{\dd x_\gamma} \right)^\text{eff}
	\approx
		\begin{cases}
			\frac{12}{\alpha^2} x_\gamma^3 & \text{for } x_\gamma \ll \alpha^2
			\\
			3 x_\gamma & \text{for } x_\gamma \gg \alpha^2.
		\end{cases}
\end{equation}
Clearly, for photons with $x_\gamma\ll\alpha^2$, the spectrum is cubic in the photon energy as required by Low's theorem. 
For photons in the energy range $\alpha^2 \ll x_\gamma \ll 1$, both
the effective theory and tree level electroweak spectra are approximately linear
with a slope of 3.

\begin{figure}[h]
	\centering
	\subfloat[]{	\label{fig:oRatio1}
			\includegraphics[width=.5\textwidth]{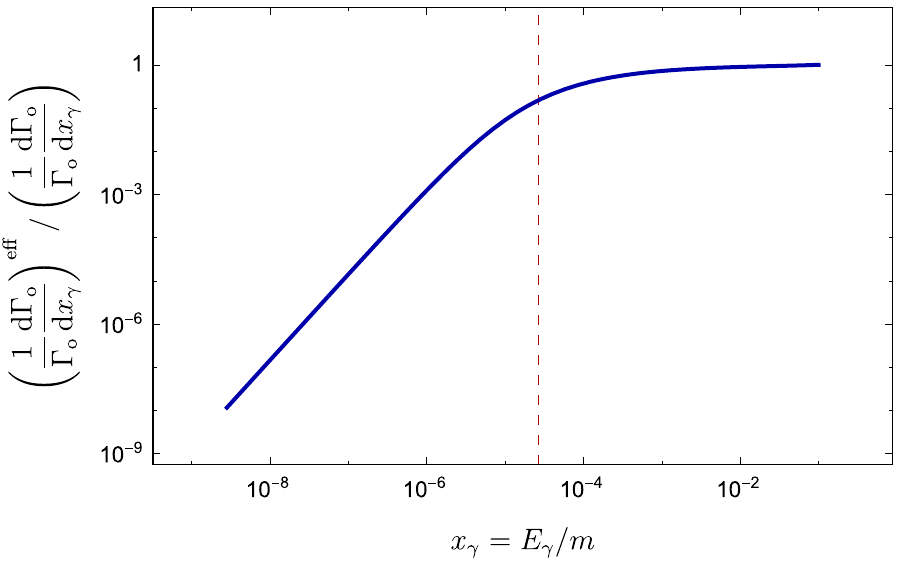}
			}
	\subfloat[]{\label{fig:oRatio2}
		\includegraphics[width=.5\textwidth]{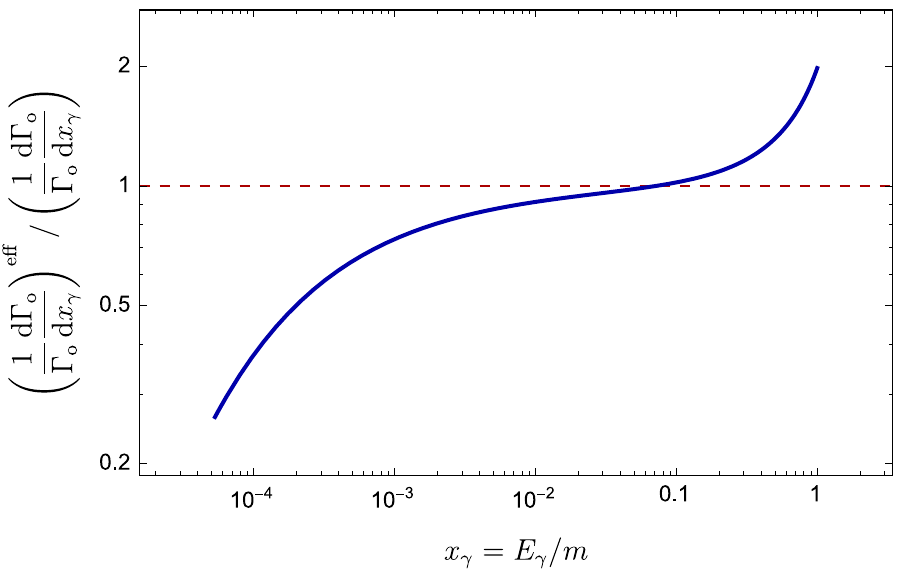}
		}
	\caption{\label{fig:oRatio}
		Log-log plot of the ratio of the effective theory amplitude to the tree 
		level electroweak amplitude for $\text{p-Ps}\to\gamma\nu\bar{\nu}$ 
		decays for \protect\subref{fig:pRatio1} the low-energy limit 
		$\alpha^4 < x_\gamma < 0.1$ and for \protect\subref{fig:pRatio2}
		the high-energy limit $\alpha^2 < x_\gamma < 1$.
		The vertical line in \protect\subref{fig:pRatio1} indicates the 
		binding energy ($x_\gamma = \alpha^2/2$) 
		while the horizontal line in \protect\subref{fig:pRatio2} is placed at 1 
		where the effective theory and electroweak theory 
		spectra are equal. 
		}
\end{figure}

The ratio of the effective theory spectrum to the tree level electroweak spectrum for $\ognn$ decays is plotted in Fig.~\ref{fig:oRatio}.
The effective theory spectrum and tree level electroweak spectrum are approximately equal in the intermediate  energy range $x_\gamma \sim \mathcal{O}(10^{-2}-10^{-1})$ (Fig.~\ref{fig:oRatio}). For high energy photons the ratio spikes upward indicating that the effective theory spectrum differs significantly from the tree level electroweak spectrum and is no longer accurate (Fig.~\protect\subref*{fig:oRatio2}).
Below the binding energy, the ratio in the log-log plot is linear with a slope of slope of 2 since the effective theory spectrum is cubic in $x_\gamma$ while the tree level electroweak spectrum is linear (Fig.~\protect\subref*{fig:oRatio1}).

\section{\label{sec:conclusions} Conclusions}

We calculated the decay rate and photon spectrum of the decay of Ps
into a photon and a neutrino-antineutrino pair ($\gnn$). Both Ps spin
states have access to the $\gamma\nu_\ell\bar{\nu}_\ell$ decay channel
where the p-Ps and o-Ps final states are orthogonal despite being
comprised of the same particles. The decay rates are given by
\eqref{eq:p-Ps rate} and \eqref{eq:o-Ps rate} and the tree 
level electroweak photon spectrum by \eqref{eq:p-Ps spectrum} 
and \eqref{eq:o-Ps spectrum}.
These rates and spectra were further examined by calculating the 
angular dependence of the decay amplitudes, angular distributions and 
spectra for specific $\gamma+Z^*$ final states 
(Tables~\ref{tab:p-Ps amp}--\ref{tab:p-Ps partial DR}). 

In principle, this decay could be observed. Experimentally, this
channel would appear as the decay of Ps into a single photon if
the neutrinos go undetected. Experimental detection of this channel
would however be very challenging given the small branching ratios.
 
The soft photon limit of the tree level electroweak spectra (equations
\eqref{eq:p-Ps spectrum} and \eqref{eq:o-Ps spectrum}) was compared
with that predicted by Low's theorem and found to be in
disagreement. This contradiction was resolved by including binding
effects in the computation of the soft photon spectrum using the
methods of non-relativistic effective field theories. The effective
theory spectra are given by equations \eqref{eq:fp Low} and
\eqref{eq:fo Low}, and are valid for photon energies much less than
the electron mass.

For photon energies much larger than the hyperfine splitting yet still
much smaller than the electron rest mass
($m\alpha^4 \ll E_\gamma \ll m$), the $\pgnn$ effective theory
spectrum approaches the tree level electroweak spectrum \eqref{eq:p-Ps
  spectrum}.  Below the hyperfine splitting
($E_\gamma \ll m\alpha^4$), the effective theory spectrum is cubic in
the soft photon energy as required by Low's theorem.  In the dipole
approximation of the Coulomb interaction, soft photon $\pgnn$ decays
proceed only by the magnetic M1 transition.

The $\ognn$ effective theory spectrum approaches the tree level
electroweak spectrum \eqref{eq:o-Ps spectrum} for photon energies much
larger than the binding energy but still much smaller than the
electron rest mass($m\alpha^2 \ll E_\gamma \ll m$).  For photon
energies much smaller than the binding energy
($E_\gamma \ll m\alpha^2$), the effective theory spectrum is cubic in
the photon energy as required by Low's theorem.  In the dipole
approximation of the Coulomb interaction, soft photon $\ognn$ decays
proceed only by the electric E1 transition.

Lastly, we find connection between the Stark effect and the soft
photon limit of the $\ognn$ spectrum and the E1 contribution to the
$\oggg$ spectrum.

\section*{Acknowledgements}
This research was supported by the Natural Sciences and Engineering Research Council of Canada (NSERC). 
We thank  Robert Szafron and Mikhail Voloshin  for helpful discussions. 
We also thank the Max Planck Institut f{\"u}r Physik, where part of this work was completed, for hospitality.

%

\section*{\label{sec:appA} Appendix A: Formulation of the $\boldsymbol{\gnn}$ decay rate in terms of $\boldsymbol{\gamma}$ and $\boldsymbol{Z^*}$}

The Feynman diagrams relevant for the $\gnn$ decay are illustrated in Fig. \ref{fig:gnn diagrams}. 
As in Sec.~\ref{sec:RandS} we neglect the 3-momentum of the incoming leptons and the virtual $W$ and $Z$ bosons.  
With these approximations, the $(\text{p/o})\text{-}\gnn$ amplitudes are
\begin{eqnarray}
	i\mathcal{M}_\text{p/o}
	& = & 
		\frac{i \GF}{\sqrt{2\pi\alpha}} (\epsilon_\gamma)^*_\mu g_{\nu \rho}
		\text{Tr}\left[X_\text{p/o}^{\mu \nu} (p_1,k_\gamma) \right]
		J^\rho (k_1,k_2),
\end{eqnarray}
where 
\begin{eqnarray}
	X_\text{p/o}^{\mu \nu} (p_1, k_\gamma)
		& = & \text{Tr} 
			\left[ 
				2m \Psi_\text{p/o} 
				\left(
					 (ie) \gamma^\nu 
					 	\left(
							b_\ell - a_\ell \gamma^5
						 \right)
			 		 \frac{\cancel{p}_1 - \cancel{k}_\gamma + m}{(p_1 - k_\gamma)^2 - m^2}  (-ie) \gamma^\mu
				\right.
			\right.
			\nonumber \\
			& & \;\;\;\;\;\;\;\;\;\;\;\;\;\;
			\left.
				\left.
 					+ (-ie) \gamma^\mu \frac{\cancel{k}_\gamma - \cancel{p}_2 + m}{(k_\gamma-p_2)^2-m^2} 
						(ie) \gamma^\nu 
				 			\left(
								b_\ell - a_\ell \gamma^5 
							\right)
				\right)	
			\right],
\end{eqnarray}
and $J^\mu \left(k_{1},k_{2}\right)=\bar{u}(k_{1})\gamma^{\mu}\left(1-\gamma^{5}\right)v(k_{2})$
is the neutral weak current. The p-Ps and o-Ps projection
operators are given by $\Psi_\text{p}=\left(1+\gamma^{0}\right)\gamma^{5}/(2\sqrt{2})$
and $\Psi_\text{o}=\left(1+\gamma^{0}\right)\boldsymbol{\gamma}\cdot\boldsymbol{\xi}/(2\sqrt{2})$
where $\boldsymbol{\xi}$ is the o-Ps polarization vector
\cite{Czarnecki:1999mw}. 

To calculate the $\gnn$ decay rate, we start from the standard formula,
\begin{eqnarray}
	\Gamma_\text{p/o} 
	& = & 
	 \frac{1}{2m_{\text{Ps}}} \int \dd \Phi_{3} (p_1+p_2;k_1,k_2,k_\gamma) 
	\frac{ \left| \psi_\text{0}(0) \right|^2}{m} \frac{1}{g} \sum_\text{spin/pol} \left| \M_\text{p/o} \right|^{2},
	\label{eq:general decay rate formula}
\end{eqnarray}
where $\psi_\text{0}(0)$ is the ground state positronium wave function at the origin and $g$ is the number of Ps polarizations of the initial state \cite{Peskin:1995ev}.

Substituting the three-body spin averaged matrix element squared 
\begin{eqnarray}
	\sum_\text{spin/pol} \left| \M_\text{p/o} \right|^2 
	& = & 
	g_{\alpha \beta} g_{\mu \rho} g_{\nu \sigma} 
	\frac{\GF^2}{2 \pi \alpha}
	X_\text{p/o}^{\alpha \mu} {X_\text{p/o}^{\beta \nu}}^*  
	\text{Tr} \left[ 
				\cancel{k}_1 \gamma^\rho (1-\gamma^{5})
					\cancel{k}_2 \gamma^\sigma (1-\gamma^{5})
			\right],
\end{eqnarray}
into \eqref{eq:general decay rate formula} and decomposing the three-body
phase space into two two-body phase spaces, yields
\begin{eqnarray}
	\Gamma_\text{p/o} 
	& = &\frac{1}{2m_\text{Ps}} \int \frac{\dd s}{2 \pi} \dd \Phi_2 (2p_1;k_\gamma,q) 
		\frac{ \left| \psi_\text{Ps}(0) \right|^2 }{m} 
		\frac{g_{\alpha \beta} g_{\mu \rho} g_{\nu \sigma} }{g}
		\frac{\GF^2}{2 \pi \alpha}
		X_\text{p/o}^{\alpha \mu} {X_\text{p/o}^{\beta \nu}}^*  
	\nonumber \\
 	&  & \int \dd \Phi_2 (q;k_1,k_2) {k_1}_\eta {k_2}_\lambda 
		\text{Tr} 
			\left[ 
				\gamma^\eta \gamma^\rho (1-\gamma^{5})
							\gamma^\lambda \gamma^\sigma (1-\gamma^{5})
			\right],
\label{eq:decay rate intermediate}
\end{eqnarray}
where $s=q\cdot q$ is the invariant mass of $Z^{*}$ squared and $q$ is its four-momentum. 
The neutrino phase space integral can be performed by writing the neutrino momentum product, ${k_1}_\eta {k_2}_\lambda$, as a linear combination of the only available tensors, ${k_1}_\eta {k_2}_\lambda =Aq^2 g_{\eta \lambda}+Bq_\eta q_\lambda$.
The momentum conserving delta function in $\dd \Phi_2(q;k_1,k_2)$ forces $q=k_{1}+k_{2}$. A system of equations for $A$ and $B$ is obtained by contracting $\int \dd \Phi_2(q;k_1,k_2) {k_1}_\eta {k_2}_\lambda $ with $g^{\eta\lambda}$ and $q^\eta q^\lambda$, and yields the solution $A=1/12$ and $B=1/6$. 
Thus, the neutrino contribution to the decay rate is 
\begin{eqnarray}
	\int \dd \Phi_2 (q;k_1,k_2) {k_1}_\eta {k_2}_\lambda 
	\text{Tr} 
		\left[ 
			\gamma^\eta \gamma^\rho (1-\gamma^{5})
						\gamma^\lambda \gamma^\sigma (1-\gamma^{5})
		\right] 
	& = & \frac{1}{3\pi}
		\left[
			q^\rho q^\sigma - q^2 g^{\rho \sigma}
		\right]
	\nonumber \\
 	& = & 
	\frac{1}{3 \pi} q^2 \sum_s
		\epsilon_s^\rho(q) {\epsilon_s^\sigma}^*(q),
\label{eq:neutrino contribution}
\end{eqnarray}
where the sum over the polarizations of a massive vector boson is
given by 
\begin{equation}
	\sum_s \epsilon_s^\rho(q) {\epsilon_s^\sigma}^*(q)
	= \frac{q^\rho q^\sigma}{q^2} - g^{\rho \sigma}.
\end{equation}
Substituting (\ref{eq:neutrino contribution}) into equation (\ref{eq:decay rate intermediate}),
we obtain the $\gnn$ decay rate in terms of $\text{Ps}\rightarrow\gamma Z^{*}$
\begin{eqnarray}
	\Gamma_\text{p/o} 
	& = & 
		\frac{1}{2m_\text{Ps}} \int \frac{\dd s}{2\pi} \dd \Phi_2(2p_1;k_\gamma,q)
		\frac{\left|\psi_{Ps}(0)\right|^{2}}{m}
		\frac{g_{\alpha \beta}}{g}
		\frac{\GF^2}{2 \pi \alpha}
		X_\text{p/o}^{\alpha \mu} {X_\text{p/o}^{\beta \nu}}^*
		\frac{1}{3\pi} q^2
		\sum_s \left(\epsilon_{s}\right)_\mu \left(\epsilon_{s}^{*}\right)_\nu
	\nonumber \\
	 & = & 
	 	\frac{\GF^2}{2 \pi^2 \alpha} \int \frac{\dd q^2}{2\pi} q^2 
		\left(
			\frac{1}{2m_\text{Ps}} 
			\int \dd \Phi_2 (2p;k_\gamma,q) 
			\frac{\left|
					\psi_{Ps}(0)
				\right|^2
				}{m}
			\frac{1}{3g} \sum_\text{pol} 
				\left| 
					\M_{\text{(p/o)-Ps} \to \gamma Z^*}
				\right|^{2}
		\right)
		\nonumber \\
 	& = & 
		\frac{\GF^2}{2 \pi^2 \alpha} \int \frac{\dd q^2}{2\pi} q^2
		\Gamma_{\text{(p/o)-Ps} \to \gamma Z^*}.
\end{eqnarray}

\section*{\label{appB} Appendix B: Derivation of the $\text{o-Ps}$ amplitudes with their angular dependencies}

Initially, the o-Ps atom is in a state of definite angular momentum denoted by $|\Lambda\rangle$. Since o-Ps and its decay products, $\gamma$ and $Z^{*}$, are all spin one particles, we abbreviate the angular momentum states $|1,m_{s}\rangle$ by $|m_{s}\rangle$ where $m_{s}$ is the projection of spin along the $z$-axis. The massive $Z^{*}$ boson has access to all three spin projection states (i.e., $m_Z \in \{\pm1,0\}$) while the massless photon cannot access the longitudinally polarized $|0\rangle$ state (i.e., $m_{\gamma} \in \{\pm1\}$). Conservation of angular momentum requires that the spin projection quantum numbers satisfy $m_{\gamma}+m_{Z}=m_{\Lambda}$; as a result, there are four different modes in which o-Ps can decay along the $z$-axis. 

Consider $|\Lambda\rangle$ initially polarized in the state $|+\rangle$ along the $z$-axis. Since the photon must have $m_{\gamma}=\pm1$, conservation of angular momentum implies $|\gamma\rangle=|+\rangle$ and $|Z^{*}\rangle=|0\rangle$; we assign the amplitude $A_{+0}$ to this decay. 
If $|\Lambda\rangle$ is initially polarized in the state $|-\rangle$, $|\gamma\rangle=|-\rangle$ and $|Z^{*}\rangle=|0\rangle$; we assign the amplitude $A_{-0}$ to this decay. 
Lastly, if $|\Lambda\rangle$ is initially polarized in the state $|0\rangle$, $m_{\gamma}=-m_{Z}$ and therefore $|\gamma\rangle=|\pm\rangle$ and $|Z^{*}\rangle=|\mp\rangle$; we assign amplitudes $A_{\pm\mp}$ to these decays. 

The $\text{o-Ps}\to\gamma Z^*$ amplitudes along the $z$-axis are 
\begin{eqnarray}
A_{\pm0} 
	& = & \pm\frac{4e^2 a_\ell}{q}
		\left[\frac{E_\gamma +E_{Z}}{q}\boldsymbol{\xi}\cdot	\boldsymbol{\epsilon}_{\pm}^{*}\right]
			= \pm\frac{e^{2}}{\sqrt{2}}\frac{E_\gamma+E_{Z}}{q}\delta_{m_{\Lambda},\pm},
\\
A_{\pm\mp} 
	& = & \pm\frac{4ie^2 a_\ell}{\sqrt{2}}\boldsymbol{\xi}\cdot\hat{{\bf z}}
		=\pm\frac{ie^{2}}{\sqrt{2}}\delta_{m_{\Lambda},0},
\end{eqnarray}   
where $\boldsymbol{\epsilon}_{\pm}$ are the transverse polarization vectors of the photon and $\boldsymbol{\xi}$ is the o-Ps polarization vector. Here $q$ is the momentum of the $Z^*$.

To determine the angular dependence of the decay amplitudes on the spherical angles, $\theta$ and $\phi$, we consider two coordinate systems $\{x,y,z\}$ and $\{x^{\prime},y^{\prime},z^{\prime}\}$. The $z^{\prime}$-axis is defined by the angles $\theta$ and $\phi$ in the $\{x,y,z\}$ coordinate system and represents the decay axis. The angular dependence of the decay amplitudes is constructed by rotating the initial o-Ps state  and then considering the decay into $\gamma+Z^*$ along $z^{\prime}$. 

\begin{figure}[h]
	\noindent 
	\centering
	\includegraphics[width=.3\textwidth]{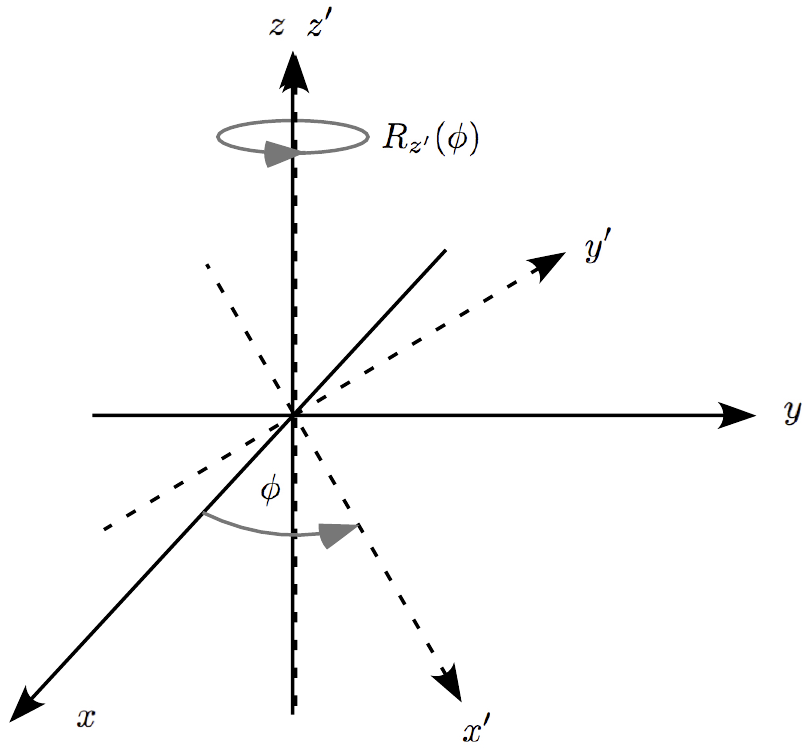}
	\includegraphics[width=.3\textwidth]{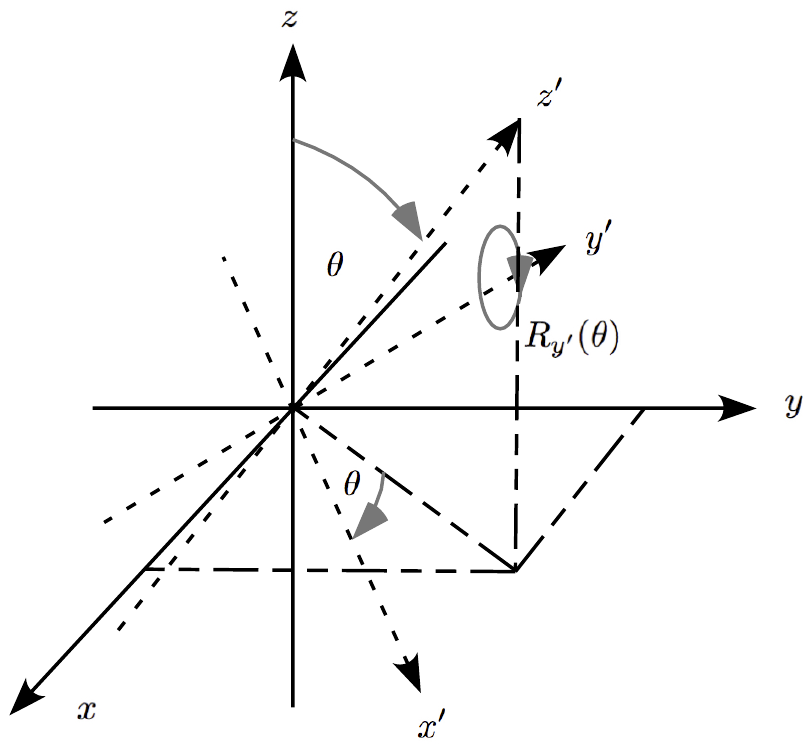}
	\includegraphics[width=.3\textwidth]{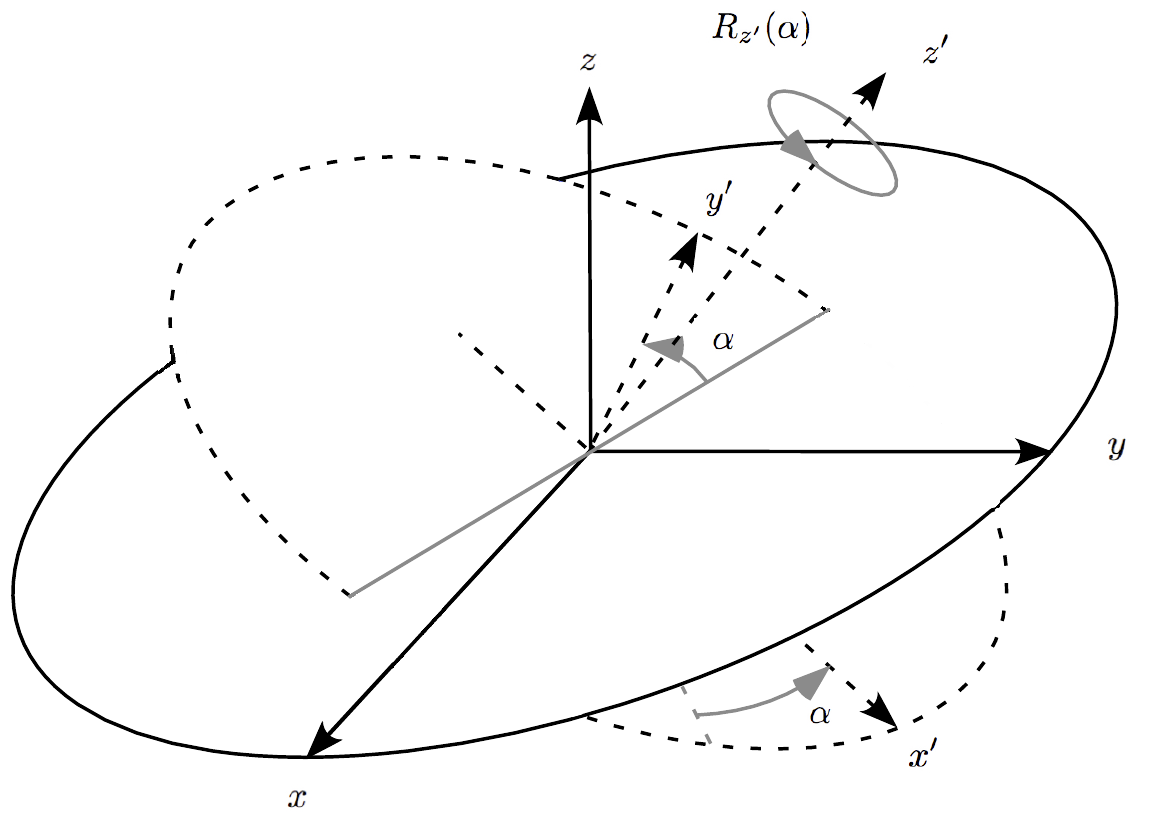}
	\caption{
		Sequence of rotations that transforms $\{x,y,z\}$ (solid) 
		to $\{x^{\prime},y^{\prime},z^{\prime}\}$ (dashed). 
		}
	\label{fig:Rotation}
\end{figure}

The combination of rotations required to bring $\{x,y,z\}$ onto $\{x^{\prime},y^{\prime},z^{\prime}\}$ (Fig. \ref{fig:Rotation}) is determined to be
\begin{equation}
	R=R_{z^{\prime}}(\alpha)R_{y^{\prime}}(\theta)R_{z^{\prime}}(\phi),
\end{equation}
where $R_{{\bf n}}(\theta)=e^{i\theta{\bf n}\cdot{\bf S}}$ is the operator for rotations about the axis given by the unit vector, $\mathbf{n}$, and ${\bf S}=(S_{x},S_{y},S_{z})$ is the spin-one matrix operator \cite{dick2016advanced}. 

Application of $R$ to $|\Lambda\rangle$ yields the amplitude for $\vert\Lambda\rangle$ to be in the state $\vert m_\Lambda^\prime \rangle$ along the $z^{\prime}$-axis for each $m_\Lambda^\prime\in\{\pm1, 0\}$. 
If $|\Lambda\rangle$ is initially polarized in the state $|+\rangle$, then $|\Lambda\rangle$ has an amplitude of $\frac{1}{2}(1+\cos\theta)e^{i\alpha}e^{i\phi}$ to be in the state $|+^{\prime}\rangle$ (the $m_{\Lambda}=1$ state along the $z^{\prime}$ axis). 
If $|\Lambda\rangle$ is in the state $|+^{\prime}\rangle$, it decays to $\vert+^{\prime};k\hat{\mathbf{z}}^{\prime}\rangle_{\gamma}\otimes\vert0^{\prime};-k\hat{\mathbf{z}}^{\prime}\rangle_{Z}$ with an amplitude $A_{+^{\prime}0^{\prime}}$, where $k$ is the magnitude of the photon momentum along $z^{\prime}$. 
Thus, the total amplitude for the decay of an o-Ps atom with spin projection $m_\Lambda =+1$ into a photon moving along $+z^{\prime}$-axis with spin projection $m^\prime_\gamma =+1$ is
\begin{equation}
	\mathcal{A}^{m_\Lambda=+}_{+^{\prime}0^{\prime}}\left(\theta,\phi\right)=\frac{A_{+^{\prime}0^{\prime}}}{2}(1+\cos\theta)e^{i\alpha}e^{i\phi}.
\end{equation}
Similarly, the amplitude for the final state $\vert-^{\prime};k\hat{\mathbf{z}}^{\prime}\rangle_{\gamma}\otimes\vert0^{\prime};-k\hat{\mathbf{z}}^{\prime}\rangle_{Z}$ is 
\begin{equation}
	\mathcal{A}^{+}_{-^{\prime}0^{\prime}}\left(\theta,\phi\right)=\frac{A_{-^{\prime}0^{\prime}}}{2}(1-\cos\theta)e^{-i\alpha}e^{i\phi},
\end{equation}
and the amplitudes for $\vert\pm^{\prime};k\hat{\mathbf{z}}^{\prime}\rangle_{\gamma}\otimes\vert\mp^{\prime};-k\hat{\mathbf{z}}^{\prime}\rangle_{Z}$
are 
\begin{equation}
	\mathcal{A}^{+}_{\pm^{\prime}\mp^{\prime}}\left(\theta,\phi\right)=\frac{-A_{\pm^{\prime}\mp^{\prime}}}{\sqrt{2}}\sin\theta e^{i\phi}.
\end{equation} 

We denote the o-Ps decay amplitudes with their full angular dependencies as $\mathcal{A}_{m_{\gamma}^{\prime}m_{Z}^{\prime}}^{m_{\Lambda}}$ where $m_{\Lambda}\in\{\pm1,0\}$ is the initial spin projection of o-Ps along the $z$-axis, and, $m_{\gamma}^{\prime}\in\{\pm1\}$ and $m_{Z}^{\prime}\in\{\pm1,0\}$ are the spin projections of the photon and $Z^{*}$ along the $z^{\prime}$-axis. 
The amplitudes, $\mathcal{A}^0_{m_\gamma ^\prime m_Z^\prime}$, are obtained using the method outlined above while $\mathcal{A}^-_{m_\gamma ^\prime m_Z^\prime}$ is obtained from $\mathcal{A}^+_{m_\gamma ^\prime m_Z^\prime}$ by the prescription $\theta\rightarrow\theta+\pi$, $\phi\rightarrow-\phi$ and $\alpha\rightarrow-\alpha$. 
The o-Ps amplitudes, $\mathcal{A}^{m_\Lambda}_{m_\gamma ^\prime m_Z^\prime}$, are listed in  table \ref{tab:o-Ps amp} where we have chosen the convention $\alpha=0$.

\section*{\label{sec:appC}Appendix C: Derivation of the $\boldsymbol{e^{+}e^{-}\rightarrow\nu\bar{\nu}}$ annihilation operator}

\begin{figure}[h]
	\subfloat[]{\label{fig:nnZ}
		\includegraphics[width=.3\textwidth]{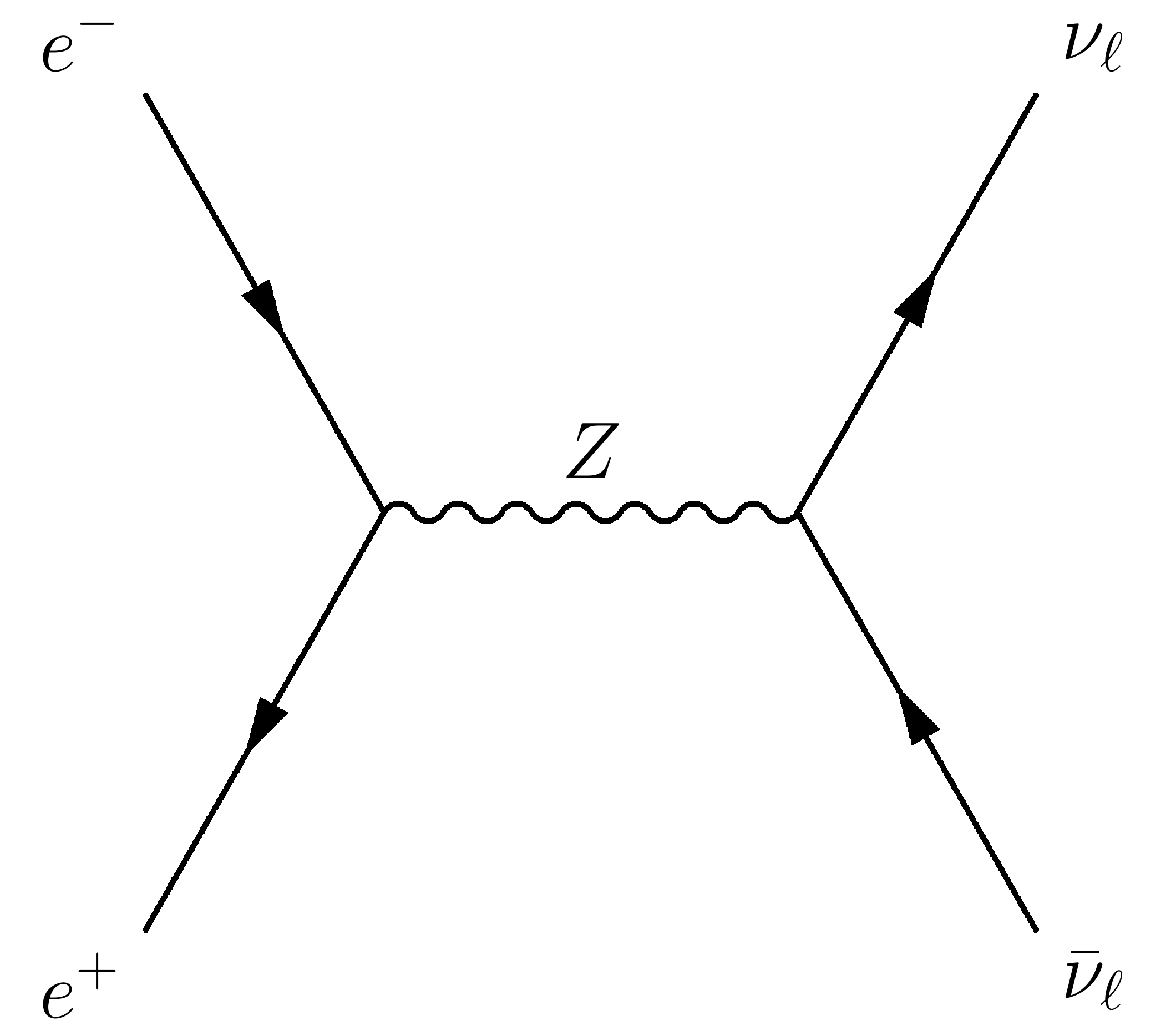}
		}
		\subfloat[]{\label{fig:nnW}
		\includegraphics[width=.3\textwidth]{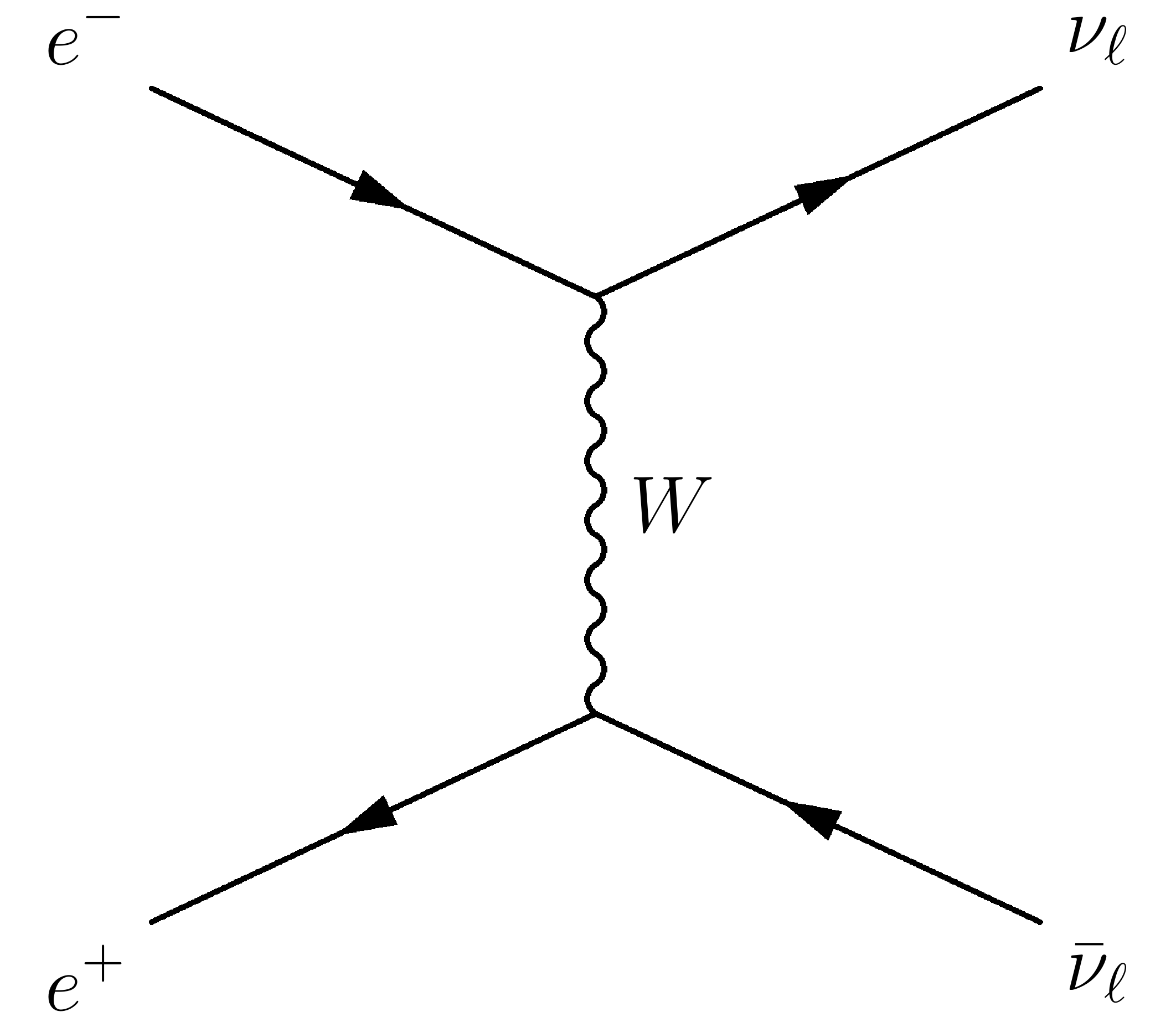}
		}
	\caption{\label{fig:nn}
	The $e^+e^- \rightarrow \nu_\ell \bar{\nu}_\ell$ annihilation graphs for
	\protect\subref{fig:nnZ} $Z$ boson exchange and 
	\protect\subref{fig:nnW} $W$ boson exchange. 
	The graph \protect\subref{fig:nnZ} contributes to 
	the $\nn$ amplitude for all lepton flavours $\ell=e,\mu,\tau$
	while \protect\subref{fig:nnW} only contributes to the amplitude when $\ell=e$. 
	}
\end{figure}

In order to calculate the effective theory amplitudes (Sec.~\ref{sec:NREFT}), we require the $\mathcal{O}(|\mathbf{p}|/m)$ expansion of the $e^+e^-\to\nu_\ell\bar{\nu}_\ell$ annihilation amplitude (Fig. \ref{fig:nn}).  The electron and positron 4-momentum are $p_1=(E,\mathbf{p})$ and $p_2=(E,-\mathbf{p})$ while the neutrino and anti-neutrino 4-momentum are $k_1$ and $k_2$. 
The amplitude of Fig. \ref{fig:nn} is
\begin{eqnarray}
	A^{(\nu_\ell \bar{\nu}_\ell)} 
	& = & - i \sqrt{2} \GF \bar{v}(-\mathbf{p}) \cancel{J} (v_\ell - a_\ell \gamma^5) u(\mathbf{p})
	\nonumber \\
	& = & 2 \sqrt{2} i \GF m  \chi^{\dagger} 
	\left[
		\left(
			\begin{array}{cc}
				\frac{\left(\boldsymbol{\sigma}\cdot\mathbf{p}\right)^{\dagger}}{2mc} & 1
			\end{array}
		\right)
		\cancel{J} \left(v_\ell-a_\ell \gamma^{5}\right) 
		\left(
			\begin{array}{c}
				1
				\\
				\frac{ \boldsymbol{\sigma} \cdot \mathbf{p} }{ 2mc }
			\end{array}
		\right)
	\right]
	\phi
	\nonumber \\
 	& = & \chi^\dagger \hat{A}^{(\nu_\ell \bar{\nu}_\ell)} \phi,
\end{eqnarray}
where 
\begin{eqnarray}
	\hat{A}^{(\nu_\ell \bar{\nu}_\ell)} & = &  2\sqrt{2} i \GF m
	\left(
		\begin{array}{cc}
			\frac{\boldsymbol{\sigma}\cdot\mathbf{p}}{2m} & 1
		\end{array}\right)
	\left(
		\begin{array}{cc}
			J_{0} & -\mathbf{J}\cdot\boldsymbol{\sigma}
			\\
			\mathbf{J}\cdot\boldsymbol{\sigma} & -J_{0}
		\end{array}
	\right)
	\left(
		\begin{array}{cc}
			v_\ell & -a_\ell
			\\
			-a_\ell & v_\ell
		\end{array}
	\right)
	\left(
		\begin{array}{c}
			1
			\\
			\frac{\boldsymbol{\sigma}\cdot\mathbf{p}}{2m}
		\end{array}
	\right)
\end{eqnarray}
is the  $\nu_\ell \bar{\nu}_\ell$ annihilation operator.
From momentum conservation, $\mathbf{k}_{1}=-\mathbf{k}_{2}$ and the time component of the neutral weak current vanishes, $J_{0}=0$.  
Therefore, the $\nu_\ell \bar{\nu}_\ell$ annihilation operator becomes
\begin{equation}
	\hat{A}^{(\nu_\ell \bar{\nu}_\ell)} 
		=  2\sqrt{2} i \GF m v_\ell \left( \mathbf{J} \cdot \boldsymbol{\sigma} \right)
			- 2\sqrt{2} \GF a_\ell \left( \mathbf{J}\times\boldsymbol{\sigma} \right) \cdot \mathbf{p}.
	\label{eq:onn ann op}
\end{equation}
The first term of equation \eqref{eq:onn ann op}, proportional to vector coupling, is the s-wave $\onn$ annihilation operator 
\begin{equation}
	\hat{A}^{(\nu_\ell \bar{\nu}_\ell)}_\text{s} = 2\sqrt{2} i \GF m v_\ell \left( \mathbf{J} \cdot \boldsymbol{\sigma} \right).
\end{equation}
In the computation of the $\pgnn$ effective theory amplitude, the s-wave annihilation operator takes the intermediate s-wave o-Ps state into a neutrino-antineutrino pair. 
The second term, proportional to axial coupling, is the p-wave $\onn$ annihilation operator
\begin{equation}
	\hat{A}^{(\nu_\ell \bar{\nu}_\ell)}_\text{p}
		= - 2\sqrt{2} \GF a_\ell \left( \mathbf{J}\times\boldsymbol{\sigma} \right) \cdot \mathbf{p}.
\end{equation}
In the computation of the $\ognn$ effective theory amplitude, the p-wave annihilation operator takes the intermediate p-wave o-Ps states into a neutrino-antineutrino pair.

\end{document}